\documentclass[1p,11pt]{elsarticle}

\journal{undisclosed}

\biboptions{numbers,sort&compress}

\usepackage{libertine}
\usepackage{libertinust1math}

\usepackage{amsmath}
\usepackage{bbold}
\usepackage{graphicx}
\usepackage{eurosym}
\usepackage{mathtools}
\usepackage{url}
\usepackage{booktabs}
\usepackage{epstopdf}
\usepackage{xfrac}
\usepackage{bm}
\usepackage[colorlinks]{hyperref}
\usepackage[nameinlink,sort&compress,capitalise,noabbrev]{cleveref}
\usepackage[leftcaption,raggedright]{sidecap}
\usepackage{subcaption}
\usepackage{blindtext}

\usepackage[parfill]{parskip}

\graphicspath{{./graphics/}}

\newcommand{\abs}[1]{\left|#1\right|}
\newcommand{\norm}[1]{\left\lVert#1\right\rVert}
\newcommand{\set}[1]{\left\{#1\right\}}
\DeclareMathOperator*{\argmin}{\arg\!\min}

\def\cA{\mathcal{A}}
\def\ba{{\bm{\alpha}}}
\def\card{\text{card}\,}
\def\x{\boldsymbol{\mathsf{x}}}

\begin{document}

\begin{frontmatter}

	\title{Broad Ranges of Investment Configurations for Renewable Power
	Systems, Robust to Cost Uncertainty and Near-Optimality}

	\author[tubaddress,kitaddress]{Fabian Neumann\corref{correspondingauthor}}
	\ead{f.neumann@tu-berlin.de}
	\author[tubaddress,kitaddress]{Tom Brown}
	\cortext[correspondingauthor]{Corresponding author}
	\address[tubaddress]{Department of Digital Transformation in Energy Systems, Institute of Energy Technology, Technische Universität Berlin (TUB), Einsteinufer 25 (TA 8), 10587, Berlin, Germany}
	\address[kitaddress]{Institute for Automation and Applied Informatics (IAI), Karlsruhe Institute of Technology (KIT), Hermann-von-Helmholtz-Platz 1, 76344, Eggenstein-Leopoldshafen, Germany}

	\begin{abstract}

To achieve ambitious greenhouse gas emission reduction targets in time, the
planning of future energy systems needs to accommodate societal preferences,
e.g.~low levels of acceptance for transmission expansion or onshore wind
turbines, and must also acknowledge the inherent uncertainties of technology
cost projections. To date, however, many capacity expansion models lean heavily
towards only minimising system cost and only studying few cost projections.
Here, we address both criticisms in unison. While taking account of technology
cost uncertainties, we apply methods from multi-objective optimisation to
explore trade-offs in a fully renewable European electricity system between
increasing system cost and extremising the use of individual technologies for
generating, storing and transmitting electricity to build robust insights about
what actions are viable within given cost ranges. We identify boundary
conditions that must be met for cost-efficiency regardless of how cost
developments will unfold; for instance, that some grid reinforcement and
long-term storage alongside a significant amount of wind capacity appear
essential. But, foremost, we reveal that near the cost-optimum a broad spectrum
of regionally and technologically diverse options exists in any case, which
allows policymakers to navigate around public acceptance issues. The analysis
requires to manage many computationally demanding scenario runs efficiently, for
which we leverage multi-fidelity surrogate modelling techniques using sparse
polynomial chaos expansions and low-discrepancy sampling.

	\end{abstract}


\end{frontmatter}

\newpage
\begin{small}
	\tableofcontents
\end{small}



\section{Introduction}
\label{sec:intro}


Energy system models have become a pivotal instrument for policy-making to find
cost-efficient system layouts that satisfy ambitious climate change mitigation
targets. But even though they have proliferated in spatial, temporal,
technological and sectoral detail and scope in recent years, least-cost
optimisation models can easily give a false sense of exactness
\cite{Trutnevyte2016, pye_modelling_2020}. Frequently, they present just a
single least-cost solution for a single set of cost assumptions, which not only
neglects uncertainties inherent to technology cost projections which capacity
expansion models are susceptible to \cite{trondle_trade-offs_2020, yue_review_2018,
pye_assessing_2018}, but also hides a wide array of alternative solutions that
are equally feasible and only marginally more expensive \cite{nearoptimal,
lombardi_policy_2020, sasse_regional_2020}.


Trade-offs revealed by deviating from least-cost solutions are extremely
attractive for policymakers, because they allow them to make decisions based on
non-economic criteria without affecting the cost-effectiveness of the system.
Knowing that many similarly costly but technologically diverse solutions exist,
helps to accommodate political and social dimensions that are otherwise hard to
quantify; for instance, rising public opposition towards reinforced transmission
lines and onshore wind turbines or an uneven distribution of new infrastructure
\cite{mccollum_energy_2020,sasse_regional_2020,schlachtberger_cost_2018}.


Techniques like multi-objective optimisation and
modelling-to-generate-alternatives are designed to find such near-optimal
solutions. Among others, they have been applied to investment planning models of
the European \cite{nearoptimal}, the Italian
\cite{lombardi_policy_2020}, and the United States power system
\cite{DeCarolis2016}, pathways to decarbonise the power system of the United
Kingdom \cite{Li2017}, a single-node sector-coupling model of Germany
\cite{nacken_integrated_2019}, global integrated assessment models
\cite{Price2017}, and were combined with a quick hull algorithm to span a
polytope of low-cost solutions for a single set of cost parameters
\cite{pedersen_modeling_2020}.


However, most of the studies above only use a central cost projection for each
considered technology. But recent decades have shown that many of these
projections contain a high level of uncertainty, particularly for fast-moving
technologies like solar, batteries and hydrogen storage. This uncertainty
propagates through the model to strongly affect the optimal and near-optimal
system compositions, thus undermining any analysis of the trade-offs. Hence, it
is crucial that apparent compromises are rigorously tested for robustness to
technology cost uncertainty to raise confidence in conclusions about viable,
cost-effective power system designs. To thoroughly sweep the uncertainty space,
we can fortunately avail of previous works on multi-dimensional global
sensitivity analysis techniques in the context of least-cost optimisation
\cite{trondle_trade-offs_2020, mavromatidis_uncertainty_2018,
pizarro-alonso_uncertainties_2019, fais_impact_2016, usher_value_2015}. We
expand their application to strengthening insights on the scope of near-optimal
trade-offs, wherein the novelty of this contribution lies.


Here, we systematically explore robust trade-offs near the cost-optimum of a
fully renewable European electricity system model, and investigate how they are
affected by uncertain technology cost projections. Thereby, we evaluate both
compromises between system cost and technology choices, as well as between pairs
of technologies. We do so by solving numerous spatially and temporally explicit
long-term investment planning problems that coordinate generation, transmission
and storage investments subject to multi-period linear optimal power flow
constraints, while employing methods from global sensitivity analysis to account
for a wide range of cost projections for wind, solar, battery and hydrogen
storage technologies.


To handle the immense computational burden incurred by searching for
near-optimal alternatives alongside evaluating many different cost parameter
sets, we employ multi-fidelity surrogate modelling techniques, based on sparse
polynomial chaos expansion that allow us to merge results from one simpler and
another more detailed model. This approach has been proven very effective in
Tröndle et al.~\cite{trondle_trade-offs_2020}. Heavy parallelisation with
high-performance computing infrastructure allowed us to solve more than 50,000
resource-intensive optimisation problems which, in combination with surrogate
modelling, admit spanning a probabilistic space of near-optimal solutions rather
than putting single scenarios into the foreground.


Thereby, we are able to present alternative solutions beyond least-cost that
have a high chance of involving a limited cost increase, just as we identify
regions that are unlikely to be cost-efficient. We derive both ranges of options
and technology-specific boundary conditions, that are not affected by cost
uncertainty and must be met to keep the total system cost within a specified
range. Our results show that indeed many such similarly costly but
technologically diverse solutions exist regardless of how technology cost
developments will unfold within the considered ranges.

\section{Methods}
\label{sec:methods}

In this section,
we first outline how we obtain least-cost and near-optimal solutions for a given cost parameter set.
We then describe the model of the European power system and define the cost uncertainties.
Finally, we explain how we make use of multi-fidelity surrogate modelling
techniques based on polynomial chaos expansions
and find an experimental design that efficiently covers the parameter space.

\subsection{Least-Cost Investment Planning}
\label{sec:leastcost}

The objective of long-term power system planning is to minimise the total
annual system costs, comprising annualised capital costs $c_\star$ for investments at locations $i$
in generator capacity $G_{i,r}$ of technology $r$, storage capacity $H_{i,s}$ of technology $s$, and transmission line capacities
$F_{\ell}$, as well as the variable operating costs $o_\star$ for generator dispatch $g_{i,r,t}$:
\begin{align}
    \min_{G,H,F,g} \quad \left\{
        \sum_{i,r}   c_{i,r}  \cdot G_{i,r}  +
        \sum_{i,s}   c_{i,s}  \cdot H_{i,s}  +
        \sum_{\ell}  c_{\ell} \cdot F_{\ell} +
        \sum_{i,r,t} w_t \cdot o_{i,r} \cdot g_{i,r,t}
    \right\}
    \label{eq:objective}
\end{align}
where the snapshots $t$ are weighted by $w_t$ such that their total duration
adds up to one year. The objective is subject to a set of linear constraints that define limits on
(i) the capacities of infrastructure from geographical and technical potentials,
(ii) the availability of variable renewable energy sources for each location and point in time, and
(iii) linearised multi-period optimal power flow (LOPF) constraints including storage consistency equations,
which we describe in more detail in the following.

The capacities of generation, storage and transmission infrastructure are
limited to their geographical potentials from above and existing infrastructure from below:
\begin{align}
    \label{eq:firstA}
    \underline{G}_{i,r}  \leq G_{i,r}  \leq \overline{G}_{i,r}  &\qquad\forall i, r \\
    \underline{H}_{i,s}  \leq H_{i,s}  \leq \overline{H}_{i,s}  &\qquad\forall i, s \\
    \underline{F}_{\ell} \leq F_{\ell} \leq \overline{F}_{\ell} &\qquad\forall \ell
\end{align}

The dispatch of a renewable generator is constrained by
its rated capacity and the time- and location-dependent availability $\overline{g}_{i,r,t}$,
given in per-unit of the generator's capacity:
\begin{align}
    0 \leq g_{i,r,t} \leq \overline{g}_{i,r,t} G_{i,r} \qquad\forall i, r, t
\end{align}
The dispatch of storage units is described by a charge variable $h_{i,s,t}^+$
and a discharge variable $h_{i,s,t}^-$, each limited by the power rating $H_{i,s}$.
\begin{align}
    0 \leq h_{i,s,t}^+ \leq H_{i,s} &\qquad\forall i, s, t \\
    0 \leq h_{i,s,t}^- \leq H_{i,s} &\qquad\forall i, s, t
\end{align}
The energy levels $e_{i,s,t}$ of all storage units are linked to the dispatch by
\begin{align}
    e_{i,s,t} =\: & \eta_{i,s,0}^{w_t} \cdot e_{i,s,t-1} + w_t \cdot h_{i,s,t}^\text{inflow} - w_t \cdot h_{i,s,t}^\text{spillage} & \quad\forall i, s, t \nonumber \\
    & + \eta_{i,s,+} \cdot w_t \cdot h_{i,s,t}^+ - \eta_{i,s,-}^{-1} \cdot w_t \cdot h_{i,s,t}^-.
\end{align}
Storage units can have a standing loss $\eta_{i,s,0}$, a charging efficiency $\eta_{i,s,+}$, a discharging efficiency $\eta_{i,s,-}$,
natural inflow $h_{i,s,t}^\text{inflow}$ and spillage $h_{i,s,t}^\text{spillage}$.
The storage energy levels are assumed to be cyclic and are constrained by their energy capacity
\begin{align}
    e_{i,s,0} = e_{i,s,T} &\qquad\forall i, s \\
    0 \leq e_{i,s,t} \leq \overline{T}_s \cdot H_{i,s} &\qquad\forall i, s, t.
\end{align}
To reduce the number of decisison variables, we link the energy capacity to
power ratings with a technology-specific parameter $\overline{T}_s$
that describes the maximum duration a storage unit can discharge at full power rating.

Kirchhoff's Current Law (KCL) requires local generators and storage units as well as
incoming or outgoing flows $f_{\ell,t}$ of incident transmission lines $\ell$
to balance the inelastic electricity demand $d_{i,t}$ at each location $i$ and snapshot $t$
\begin{align}
    \sum_r g_{i,r,t} + \sum_s h_{i,s,t} + \sum_\ell K_{i\ell} f_{\ell,t} = d_{i,t} \qquad\forall i,t,
\end{align}
where $K_{i\ell}$ is the incidence matrix of the network.

Kichhoff's Voltage Law (KVL) imposes further constraints on the flow of AC lines.
Using linearised load flow assumptions, the voltage angle difference around every closed cycle in the
network must add up to zero. We formulate this constraint using a cycle basis $C_{\ell c}$
of the network graph where the independent cycles $c$ are expressed as
directed linear combinations of lines $\ell$ \cite{cycleflows}.
This leads to the constraints
\begin{align}
    \sum_\ell C_{\ell c} \cdot x_\ell \cdot f_{\ell,t} = 0 \qquad\forall c,t
    \label{eq:kvl}
\end{align}
where $x_\ell$ is the series inductive reactance of line $\ell$.
Controllable HVDC links are not affected by this constraint.

Finally, all line flows $f_{\ell,t}$ must be operated within their nominal capacities $F_\ell$
\begin{align}
    \abs{f_{\ell,t}} \leq \overline{f}_{\ell} F_{\ell} & \qquad\forall \ell, t,
    \label{eq:lastA}
\end{align}
where $\overline{f}_\ell$ acts as a per-unit buffer capacity
to protect against the outage of single circuits.

This problem is implemented in the open-source tool PyPSA \cite{pypsa} and is solved by Gurobi.
Note, that it assumes perfect foresight for a single reference year based on which capacities are optimised.
It does not include pathway optimisation, nor aspects of reserve power, or system stability.
Changes of line expansion to line impedance are ignored.

\subsection{Near-Optimal Alternatives}
\label{sec:nearoptimal}

Using the least-cost solution as an anchor, we use the
$\epsilon$-constraint method from multi-objective optimisation
to find near-optimal feasible solutions \cite{nearoptimal,mavrotas_effective_2009}.
For notational brevity, let $c^\top x$ denote the linear objective function \cref{eq:objective}
and $Ax\leq b$ the set of linear constraints \crefrange{eq:firstA}{eq:lastA}
in a space of continuous variables,
such that the minimised system cost can be represented by
\begin{equation}
    C = \min_x\left\{c^\top x \mid Ax\leq b\right\}.
\end{equation}


We then encode the original objective as a constraint
such that the cost increase is limited to a given $\epsilon$.
In other words, the feasible space is cut to solutions that
are at most $\epsilon$ more expensive than the least-cost solution.
Given this slack, we can formulate alternative search directions in the objective.
For instance, we can seek to minimise the sum of solar installations $x_s \subseteq x$ with
\begin{equation}
    \overline{x_s} = \min_{x_s}\left\{\: 1^\top x_s \mid Ax\leq b,\quad c^\top x\leq (1+\epsilon)\cdot C \:\right\}.
\end{equation}
To draw a full picture of the boundaries of the near-optimal feasible space,
we systematically explore the extremes of various technologies:
we both minimise and maximise the system-wide investments in
solar, onshore wind, offshore wind, any wind, hydrogen storage, and battery storage
capacities, as well as the total volume of transmission network expansion.
Evaluating each of these technology groups for
different cost deviations $\epsilon \in \{1\%,2\%,4\%,6\%,8\%\}$
allows us to observe how the degree of freedom regarding investment decisions
rises as the optimality tolerance is increased, both at lower and upper ends.
The boundaries delineate Pareto frontiers on which no criterion,
neither reducing system cost nor extremising the capacity of a technology,
can be improved without depressing the other.
By arguments of convexity, these extremes even define limits
within which all near-optimal solutions are contained.
Moreover, although this scheme primarily studies aggregated capacities,
the solutions are spatially explicit, and we can inspect for each case
how the capacities of each technology are distributed within the network.


The near-optimal analysis above only explores the extremes of one technology at
a time, i.e.~one direction in the feasible space. But actually the space of
attainable solutions within $\epsilon$ of the cost-optimum is multi-dimensional.
To further investigate trade-offs between multiple technologies, in addition to the
$\epsilon$-constraint and the objective to extremise capacities of a particular
technology, we formulate a constraint that fixes the capacity of another
technology. For instance, we search for the minimum amount of wind capacity $x_w
\subseteq x$ given that a certain amount of solar is built
\begin{equation}
    \overline{x_w} = \min_{x_w}\left\{\:1^\top x_w \mid Ax\leq b,\quad c^\top x\leq (1+\epsilon)\cdot C, \quad 1^\top x_s = \underline{x_s} + \alpha \cdot (\overline{x_s}-\underline{x_s}) \:\right\}.
\end{equation}
The $\alpha$ denotes the relative position within the near-optimal
range of solar capacities at given $\epsilon$.
For example, at $\alpha=0\%$ we look for the least wind capacity
given that minimal solar capacities are built.
An alternative but more complex approach to spanning the space of near-optimal solutions in multiple dimensions at a time
using a quick hull algorithm was presented by Pedersen et al.~\cite{pedersen_modeling_2020}.


Due to computational constraints, we focus on technologies which are assumed to
lend themselves to substitution and limit the corresponding analysis to a single
cost increase level of $\epsilon=6\%$. We consider the three pairs, (i) wind and
solar, (ii) offshore and onshore wind, (iii) hydrogen and battery storage, by
minimising and maximising the former while fixing the latter at positions
$\alpha \in \{0\%,25\%,50\%,75\%,100\%\}$ within the respective near-optimal
range.

\subsection{Model Inputs}
\label{sec:inputs}

\begin{SCfigure}
        \begin{tabular}{cc}
            \footnotesize (a) low-fidelity: & \footnotesize (b) high-fidelity: \\
            \footnotesize 37 nodes and 4-hourly & \footnotesize 128 nodes and 2-hourly \\
            \includegraphics[width=0.31\textwidth]{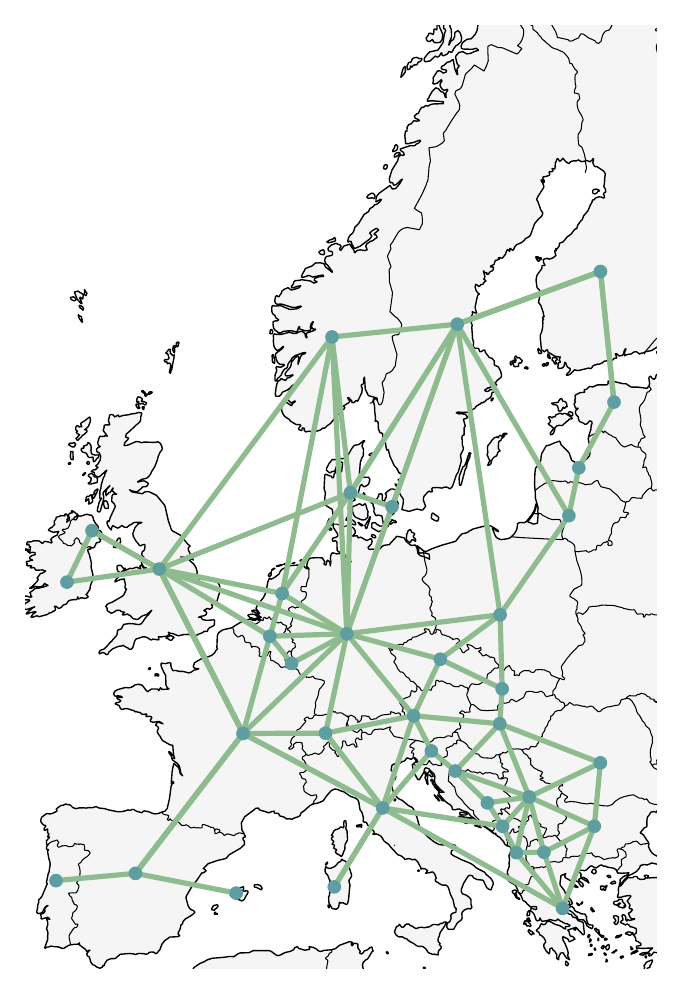} &
            \includegraphics[width=0.31\textwidth]{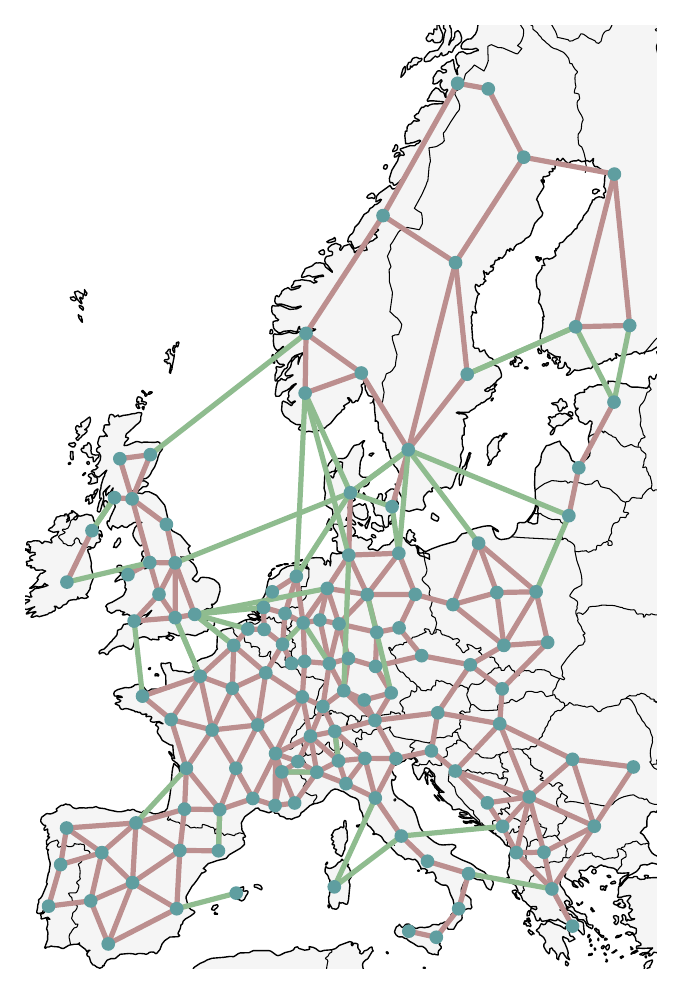} \\
            \includegraphics[width=0.31\textwidth]{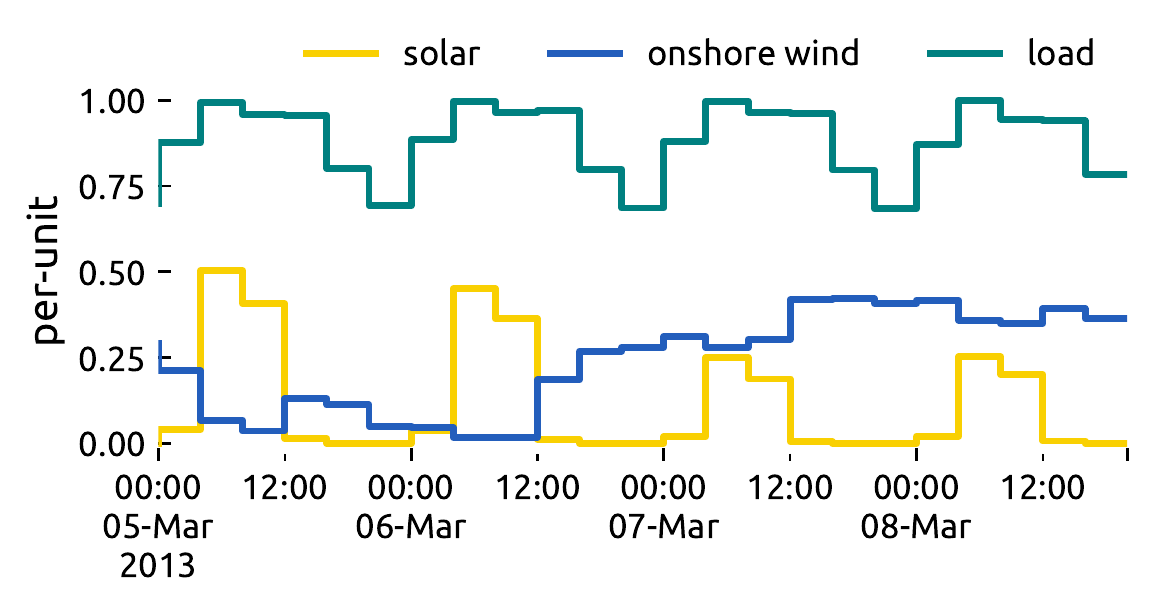} &
            \includegraphics[width=0.31\textwidth]{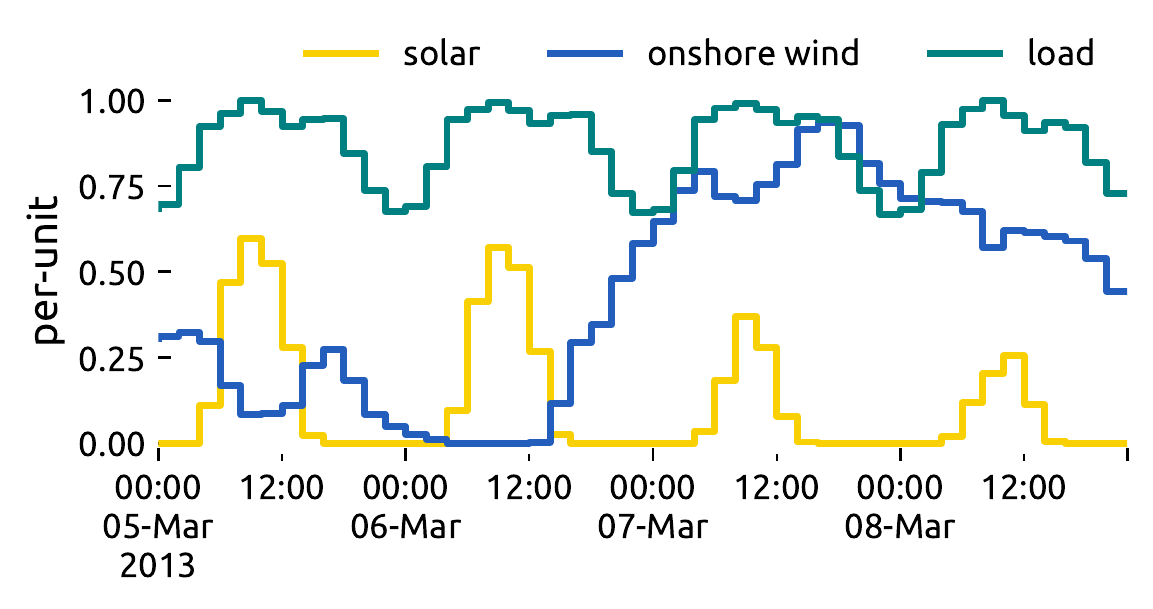} \\
        \end{tabular}
    \caption[Spatial and temporal resolution of the low and high fidelity model]{Spatial and temporal resolution of the low and high fidelity model.
    Green lines represent controllable HVDC lines. Red lines represent HVAC lines.
    The exemplary capacity factors for wind and solar are shown for four days in March
    at the northernmost node in Germany, alongside the normalised load profile.}
    \label{fig:pypsaeur}
\end{SCfigure}


The instances of the coordinated capacity expansion problem
 (\cref{sec:leastcost}) are based on \mbox{PyPSA-Eur}, which is an open model of
 the European power transmission system that combines high spatial and temporal
 resolution \cite{pypsaeur}. Because it only uses open data and every processing
 step is defined in a workflow \cite{snakemake}, we achieve a high level of
 transparency and reproducibility. In the following, we outline the main
 features and configurations, and refer to the supplementary material and Hörsch
 et al.~\cite{pypsaeur} for more details.

\paragraph{Scenario}
We target a fully renewable electricity system based on variable resources such
as solar photovoltaics, onshore wind and offshore wind, that has not carbon
emissions. We pursue a greenfield approach subject to a few notable exceptions.
The existing hydro-electric infrastructure (run-of-river, hydro dams,
pumped-storage) is included but not considered to be extendable due to assumed
geographical constraints. Furthermore, the existing transmission infrastructure
can only be reinforced continuously but may not be removed. In addition to
balancing renewables in space with transmission networks, the model includes
storage options at each node to balance renewables in time. We consider two extendable
storage technologies: battery storage representing short-term storage suited to
balancing daily fluctuations and hydrogen storage which exemplifies
long-term synoptic and seasonal storage.

\paragraph{Spatial and Temporal Resolution}
Since the spatial and temporal resolution strongly affects the size of the
optimisation problem, running the model at full resolution is computationally
infeasible. Throughout the paper, we will therefore make use of two levels of
aggregation, reflecting a compromise between the computational burden incurred
by high-resolution models and the growing inaccuracies regarding transmission
bottlenecks and resource distribution in low-resolution models. We consider a
low-fidelity model with 37 nodes at a 4-hourly resolution for a full year that
models power flow via a transport model (i.e.~excluding KVL of \cref{eq:kvl})
and a high-fidelity model with 128 nodes at a 2-hourly resolution that is
subject to linearised load flow constraints (\cref{fig:pypsaeur}).


\paragraph{Transmission Grid and Hydro-Electricity}
The topology of the European transmission network is retrieved from the ENTSO-E
transparency map and includes all lines at and above 220 kV. Capacities and
electrical characteristics of transmission lines and substations are inferred
from standard types for each voltage level, before they are transformed to a
uniform voltage level. For each line, $N-1$ security is approximated by limiting
the line loading to 70\% of its nominal rating. The dataset further includes
existing high-voltage direct current (HVDC) links and planned projects from the
Ten Year Network Development Plan (TYNDP). Existing run-of-river, hydro-electric
dams, pumped-hydro storage plants are retrieved from
\textit{powerplantmatching}, a merged dataset of conventional power plants.


\paragraph{Renewable Energy Potentials}
Eligible areas for developing renewable infrastructure are calculated
per technology and the grid nodes' Voronoi cells, assuming wind and solar installations always connect to the closest substation.
How much wind and solar capacity may be built at a location is constrained by
eligible codes of the CORINE land use database and is further restricted by distance criteria,
allowed deployment density, and the natural protection areas specified in the NATURA 2000 dataset.
Moreover, offshore wind farms may not be developed at sea depths exceeding 50 metres,
as indicated by the GEBCO bathymetry dataset.


\paragraph{Renewables and Demand Time Series}
The location-dependent renewables availability time series are generated
based on two historical weather datasets from the year 2013.
We retrieve wind speeds, run-off and surface roughness from the ERA5 reanalysis dataset and
use the satellite-aided SARAH-2 dataset for the direct and diffuse surface solar irradiance.
Models for wind turbines, solar panels, and the inflow into the basins of hydro-electric dams
convert the weather data to hourly capacity factors and aggregate these to each grid node.
Historical country-level load time series are taken from ENTSO-E statistics and are
heuristically distributed to each grid node to 40\% by population density and to 60\% by gross domestic product.


\subsection{Technology Cost Uncertainty}
\label{sec:uncertainty}

\begin{SCtable}
    \begin{small}
        \begin{tabular}{cccc}
            \toprule
            Technology & Lower Annuity & Upper Annuity & Unit  \\ \midrule
            Onshore Wind & 73 & 109 & EUR/kW/a \\
            Offshore Wind & 178 & 245 & EUR/kW/a \\ 
            Solar & 36 & 53 & EUR/kW/a \\
            Battery & 30 & 125 & EUR/kW/a \\
            Hydrogen & 111 & 259 & EUR/kW/a \\ \bottomrule
        \end{tabular}
    \end{small}
    \caption[Technology Cost Uncertainty]{Technology cost uncertainty using optimistic and pessimistic assumptions from the Danish Energy Agency \cite{DEA}.}
    \label{tab:costuncertainty}
\end{SCtable}


Uncertainty of technology cost projections is driven by two main factors:
unknown learning rates (i.e.~how quickly costs fall as more capacity is built) and
unclear deployment rates (i.e.~how much capacity will be built in the future) \cite{gritsevskyi_modeling_2000,yeh_review_2012}.
As modelling technological learning endogeneously is computationally challenging due to the nonconvexity it entails \cite{heuberger_power_2017,mattsson_learning_2019},
technology cost uncertainty is typically defined exogenously by an interval within which costs may vary
and a distribution that specifies which segments are more probable.


Ranges of cost projections are best chosen as wide as possible to avoid
excluding any plausible scenarios
\cite{moret_characterization_2017,mccollum_energy_2020}. When uncertainty has
been considered in the literature, cost assumptions have commonly been modelled
to vary between $\pm20\%$ and $\pm65\%$ depending on the technology's maturity
\cite{moret_characterization_2017,shirizadeh_how_2019,pizarro-alonso_uncertainties_2019,li_using_2020,trondle_trade-offs_2020}.
In this study, we consider uncertainty regarding the annuities of onshore wind,
offshore wind, solar PV, battery and hydrogen storage systems. The latter
comprises the cost of electrolysis, cavern storage, and fuel cells. For solar PV
we assume an even split between utility-scale PV and residential rooftop PV.
Evaluating uncertainties based on annuities has a distinct advantage. They can
be seen to simultaneously incorporate uncertainties about the overnight
investments, fixed operation and maintenance costs, their lifetime, and the
discount rate, since multiple combinations lead to the same annuity. We built
the uncertainty ranges presented in \cref{tab:costuncertainty} from the
optimistic and pessimistic technology cost and lifetime projections for the year
2050 from the Danish Energy Agency, which correspond to 90\% confidence
intervals \cite{DEA}. In cases where no uncertainty ranges were provided
for the year 2050, such as for rooftop PV, projections for the year 2030
define the upper end of the uncertainty interval.


Distributions of cost projections have been assumed to follow normal
\cite{mavromatidis_uncertainty_2018} or triangular distributions
\cite{li_using_2020}. But independent uniform distributions are the most
prevalent assumption
\cite{moret_characterization_2017,moret_robust_2016,shirizadeh_how_2019,trondle_trade-offs_2020,pilpola_analyzing_2020,Li2017,Trutnevyte2013,lopion_cost_2019}.
This approach is backed by the maximum entropy approach
\cite{trondle_trade-offs_2020}, which states that given the persistent lack of knowledge
about the distribution the independent uniform distribution, that makes fewest
assumptions, is most appropriate. Although the assumed independence may neglect
synergies between technologies, for example, between offshore and onshore wind
turbine development, we follow the literature by assuming that the cost are
independent and uniformly distributed within the ranges specified in
\cref{tab:costuncertainty}.

\subsection{Surrogate Modelling with Polynomial Chaos Expansion}
\label{sec:surrogate}


Searching for least-cost solutions (\cref{sec:leastcost}) and many associated near-optimal alternatives (\cref{sec:nearoptimal}) of a highly resolved power system model (\cref{sec:inputs}) on its own
is already labour-intensive from a computational perspective.
Repeating this search for a large variety of cost assumptions (\cref{sec:uncertainty}),
to be able to make statements about the robustness of
investment flexibility near the optimum under uncertainty,
adds another layer to the computational burden.


Surrogate models\footnote{Surrogate names are also known by names such as approximation models, response surface methods, metamodels and emulators.}
offer a solution for such cases, where the outcome of the
original model cannot be obtained easily.
In contrast to the full model, they only imitate the input/output behaviour for a selection of aggregated outputs, but take much less time to compute \cite{palar_multi-fidelity_2016}.
Like other machine learning techniques, they generalise from a
training dataset that comprises only a limited number of samples.
As surrogate models interpolate gaps in the parameter space that are not contained in the sample set,
which would otherwise be computationally expensive to fill,
they are well suited to use cases such as parameter space exploration and sensitivity analysis.


Consequently, in this paper we will make use of surrogate models that map the
cost of onshore wind, offshore wind, solar, hydrogen, and battery storage (\cref{tab:costuncertainty})
onto a selection of eight system-level outputs.
These are the total system cost and the installed onshore wind, offshore wind, solar, hydrogen, battery, and transmission network capacities.
We construct surrogate models for least-cost and near-optimal solutions separately
for each system cost slack, search direction, fixed total capacity, and output variable. This results in a collection of 808 individual surrogate models based on 101 solved optimisation problems
per set of cost assumptions.
The method we choose from an abundance of alternatives is based on polynomial chaos expansion (PCE)
\cite{sudret_global_2008,fajraoui_optimal_2017,gratiet_metamodel-based_2015}.
We select this approach because the resulting approximations
allow efficient analytical statistical evaluation \cite{sudret_global_2008} and
can conveniently combine training data from variously detailed models \cite{palar_multi-fidelity_2016}.


The general idea of surrogate models based on PCE is to
represent uncertain model outputs as a linear combination of orthogonal basis functions
of the random input variables weighted by deterministic coefficients \cite{muhlpfordt_uncertainty_2020}.
It is a Hilbert space technique that works in principle analogously to decomposing a periodic signal into its Fourier components \cite{muhlpfordt_uncertainty_2020}.
Building the surrogate model consists of the following steps:
(i) sampling a set of cost projections from the parameter space,
(ii) solving the least-cost or near-optimal investment planning problem for each sample,
(iii) selecting an expansion of orthogonal polynomials within the parameter space,
(iv) performing a regression to calculate the polynomial coefficients, and ultimately
(v) using the model approximation for statistical analysis.
In the following, we will formalise this approach mathematically,
which we implemented using the \textit{chaospy} toolbox \cite{feinberg_chaospy_2015},
and elaborate on individual aspects in more detail.

We start by defining the vector of random input variables as
\begin{equation}
    \x = \{\mathsf{x}_1, \dots , \mathsf{x}_m\}
\end{equation}
that represents the $m$ uncertain cost projections.
Further, we let
\begin{equation}
    \mathsf{y} = f(\x)
\end{equation}
describe how the uncertainty of inputs $\x$ propagates
through the computationally intensive model $f$
(i.e.~the solving a large optimisation problem)
to the outputs $\mathsf{y} \in \mathbb{R}$.

We can represent the computational model $f$ with its polynomial chaos expansion
\begin{equation}
    \mathsf{y} = f(\x) = \sum_{\ba \,\in\, \mathbb{N}^m} r_{\ba} \psi_{\ba}(\x),
    \label{eq:pce}
\end{equation}
where $\psi_\ba$ denotes multivariate orthogonal polynomials that form
a Hilbertian basis and $r_\ba \in \mathbb{R}$ are the corresponding polynomial coefficients \cite{sudret_global_2008}. The multiindex $\ba = \{\alpha_1,\dots,\alpha_m\}$
denotes the degree of the polynomial $\psi_\ba$ in each of the $m$ random input variables $\mathsf{x}_i$.
As \cref{eq:pce} features an infinite number of unknown coefficients,
it is common practice to approximate by truncating the expansion to get a finite number of coefficients
\begin{equation}
    f(\x) \approx f'(\x) = \sum_{\ba \,\in\, \cA^{m,p}} r_\ba \psi_\ba(\x).
\end{equation}
In the standard truncation scheme \cite{gratiet_metamodel-based_2015,sudret_global_2008},
all polynomials in $m$ input variables where the total degree is less than $p$ are selected.
We can write this as a set of indices
\begin{equation}
    \cA^{m,p} = \left\{\ba \in \mathbb{N}^m \,:\, \abs{\ba} \leq p\right\},
\end{equation}
where $\abs{\ba} = \sum_{i=1}^m \alpha_i$.
Given the joint distribution of $\x$ and a maximum degree,
a suitable collection of orthogonal polynomials can be constructed
using a three terms recurrence algorithm \cite{feinberg_chaospy_2015}.
The cardinality of the truncated PCE,
\begin{equation}
    q = \card \cA^{m,p} = \left(\begin{matrix}
        m+p \\
        p
    \end{matrix}\right) = \frac{(m+p)!}{m!p!},
    \label{eq:cardinality}
\end{equation}
indicates the number of unknown polynomial coefficients.

We determine these coefficicients by a regression based on
a set of cost parameter samples
and the corresponding outputs,
\begin{equation}
    \mathcal{X} = \set{ \bm x^{(1)},\dots,\bm x^{(n)} } \quad\text{and}\quad
    \mathcal{Y} = \set{ f\left(\bm x^{(1)}\right),\dots,f\left(\bm x^{(n)}\right) }.
\end{equation}
Using this training dataset, we minimise the least-square residual of the polynomial approximation across all observations.
We add an extra $L_1$ regularisation term, 
that induces a preference for fewer non-zero coefficients, and solve
\begin{equation}
    \hat{\bm{r}} = \argmin_{\bm{r} \,\in\, \mathbb{R}^q} \left[ \frac{1}{n} \sum_{i=1}^n \left(
        f\left(\bm x^{(i)}\right) - \sum_{\ba \,\in\, \cA^{m,p}} r_\ba \psi_\ba\left(\bm x^{(i)}\right)
        \right)^2  + \lambda \norm{\bm{r}}_1 \right],
        \label{eq:regression}
    \end{equation}
where we set the regularisation penalty to $\lambda=0.005$.
This results in a sparse PCE that has proven to
improve approximations in high-dimensional uncertainty spaces
and to reduce the required number of samples for comparable approximation errors \cite{gratiet_metamodel-based_2015}.
Knowing the optimised regression coefficients, we can now assemble the complete surrogate model
\begin{equation}
    \mathsf{y} = f(\x) \approx f'(\x) = \sum_{\ba \,\in\, \cA^{m,p}} \hat{r}_\ba \psi_\ba (\x).
\end{equation}

\subsection{Multifidelity Approach}
\label{sec:multifidelity}


To construct a sufficiently precise PCE-based surrogate model, it is desirable
to base it on many samples from a high-fidelity model. However, this is likely
prohibitively time-consuming. On the other hand, relying only on samples from a
low-fidelity model may be too inaccurate \cite{ng_multifidelity_2012}. For
example, an investment model that features only a single node per country will
underestimate transmission bottlenecks and regionally uneven resource or demand
distribution. In \cref{sec:inputs} we already alluded to using two models with
varying spatial and temporal resolution in this paper. We integrate both in a
multi-fidelity approach \cite{ng_multifidelity_2012,palar_multi-fidelity_2016},
and demonstrate how we can simultaneously avail of high coverage of the
parameter space by sampling the simpler model many times, and the high
spatio-temporal detail yielded by fewer more complex model runs.


The idea of the multi-fidelity approach is to build a corrective surrogate model $\Delta'(\x)$ for the
error of the low-fidelity model $f_\ell$ compared to the high-fidelity model $f_h$
\begin{equation}
    \Delta(\x) = f_h(\x) - f_\ell(\x),
\end{equation}
and add it to a surrogate model of the low-fidelity model to approximate the behaviour of the
high-fidelity model
\begin{equation}
    f_h'(\x) = f_\ell'(\x) + \Delta'(\x).
\end{equation}
Typically, the corrective PCE rectifies only the lower order effects
of the low-fidelity surrogate model \cite{palar_multi-fidelity_2016}.
The advantage is that this way the correction function can be determined
based on fewer samples analogous to \cref{sec:surrogate}.
To sample the errors, it is only required that the
high-fidelity samples are a subset of the low-fidelity samples, e.g.
\begin{equation}
    \mathcal{X}_h = \set{ \bm{x}^{(1)}, \dots, \bm{x}^{(n_h)}} \quad\text{and}\quad
    \mathcal{X}_\ell = \set{ \bm{x}^{(1)}, \dots, \bm{x}^{(n_h)}, \dots, \bm{x}^{(n_\ell)}},
\end{equation}
which we can easily guarantee by using deterministic low-discrepancy series
in the experimental design (\cref{sec:sampling}).
With $p_c < p_\ell$ and consequently $\cA_c \subset \cA_\ell$,
the multi-fidelity surrogate model can be written as a combination of low-fidelity
and corrective polynomial coefficients
\begin{equation}
    f_h' (\x) = \sum_{\ba\,\in\,\cA_\ell^{m,p_\ell} \,\cap\, \cA_c^{m,p_c}}
    \left(
     r_{\ell,\ba} + r_{c,\ba}
    \right) \psi_\ba(\x) +
    \sum_{\ba\,\in\,\cA_\ell^{m,p_\ell} \,\setminus\, \cA_c^{m,p_c}}
    r_{\ell,\ba} \psi_\ba(\x).
\end{equation}


In this work, we apply a multi-fidelity surrogate model that considers
effects up to order three observed in the low-fidelity model. These are then corrected with linear terms derived from insights from the high-fidelity model.
We justify this choice by experimentation in \cref{sec:validation},
by testing against other typical choices between orders one to five \cite{gratiet_metamodel-based_2015}.
Given the polynomial expansion order, the remaining question is how many samples are necessary to attain an acceptable approximation.

\subsection{Experimental Design}
\label{sec:sampling}



The experimental design covers strategies to find sufficiently high coverage
of the parameter space at low computational cost \cite{fajraoui_optimal_2017,usher_value_2015}.
It deals with how many samples are drawn and what sampling method is used.


Traditional Monte-Carlo sampling with pseudo-random numbers is known to
possess slow convergence properties, 
especially in high-dimensional parameter spaces.
So-called low-discrepancy series can greatly improve on random sampling.
Because they are designed to avoid forming large gaps and clusters,
these deterministic sequences efficiently sample from the parameter space \cite{fajraoui_optimal_2017}.
Thus, we choose to draw our samples from a low-discrepancy Halton sequence.


For the question about how many samples should be drawn,
we resort to the oversampling ratio (OSR) as a guideline.
The OSR is defined as the ratio between the number of samples
and the number of unknown coefficients \cite{palar_multi-fidelity_2016}.
The literature recommends values between two and three \cite{hosder2007,palar_multi-fidelity_2016,fajraoui_optimal_2017,gratiet_metamodel-based_2015}.
In other words, for a sufficiently accurate approximation,
there should be significantly more samples than unknown coefficients.
If the OSR is lower, the regression is prone to the risk of overfitting.
On the other hand, a high OSR may lead to a very coarse approximation \cite{palar_multi-fidelity_2016}.

According to \cref{eq:cardinality}, targeting an OSR of two
and considering the five uncertain technology cost parameters (\cref{tab:costuncertainty}),
approximating linear effects would require at least 12 samples, whereas cubic relations
would already need 112 samples. Even 504 samples would be necessary to model the dynamics of order 5.
To investigate the quality of different PCE orders
and retain a validation dataset,
we draw 500 samples for the low-fidelity model.
Due to the computational burden carried by the high-fidelity models,
we settle on a linear correction in advance, such that
15 samples for the high-fidelity model are acceptable.
In combination with 101 least-cost and near-optimal optimisation
runs for each sample, this setup results in a total number of
50,500 runs of the low-fidelity model and
1,515 runs of the high-fidelity model.
On average a single high-fidelity model run took 20 GB of memory and 5 hours to solve.
Each low-fidelity model run on average consumed 3 GB of memory and completed within 5 minutes.
This setup profits tremendously from parallelisation as it involves
numerous independent optimisation runs.
Moreover, it would have been infeasible to carry out without high-performance computing.

\subsection{Model Validation}
\label{sec:validation}

\begin{figure}
    \noindent\makebox[\textwidth]{
        \begin{subfigure}[t]{1.4\textwidth}
            \caption{number of samples}
            \label{fig:error:samples}
            \includegraphics[height=.2\textwidth, trim=0cm 0cm 4.5cm 0cm, clip]{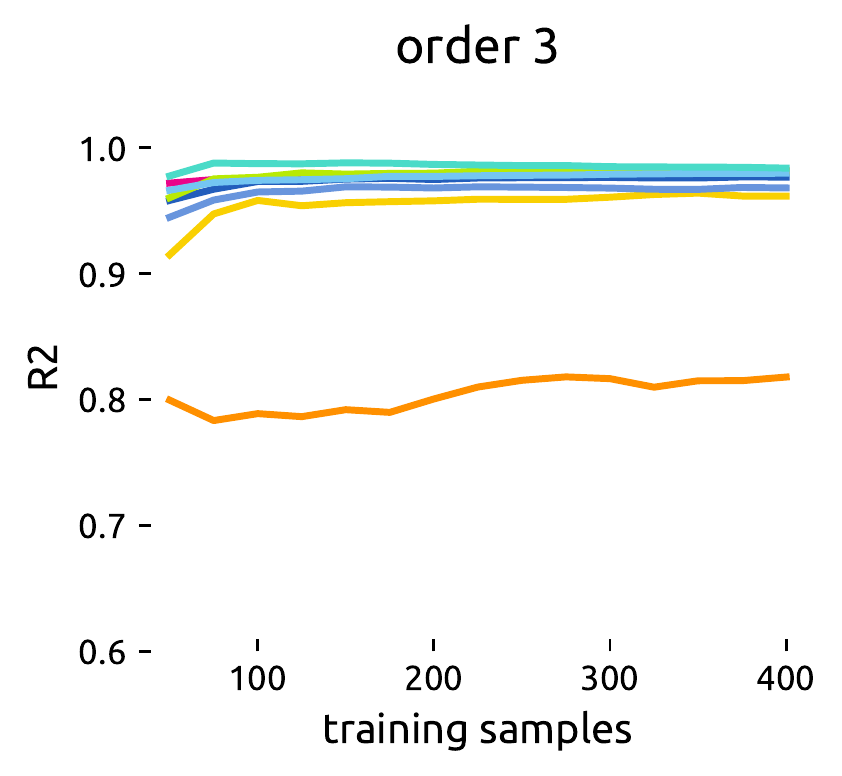}
            \includegraphics[height=.2\textwidth, trim=0cm 0cm 4.5cm 0cm, clip]{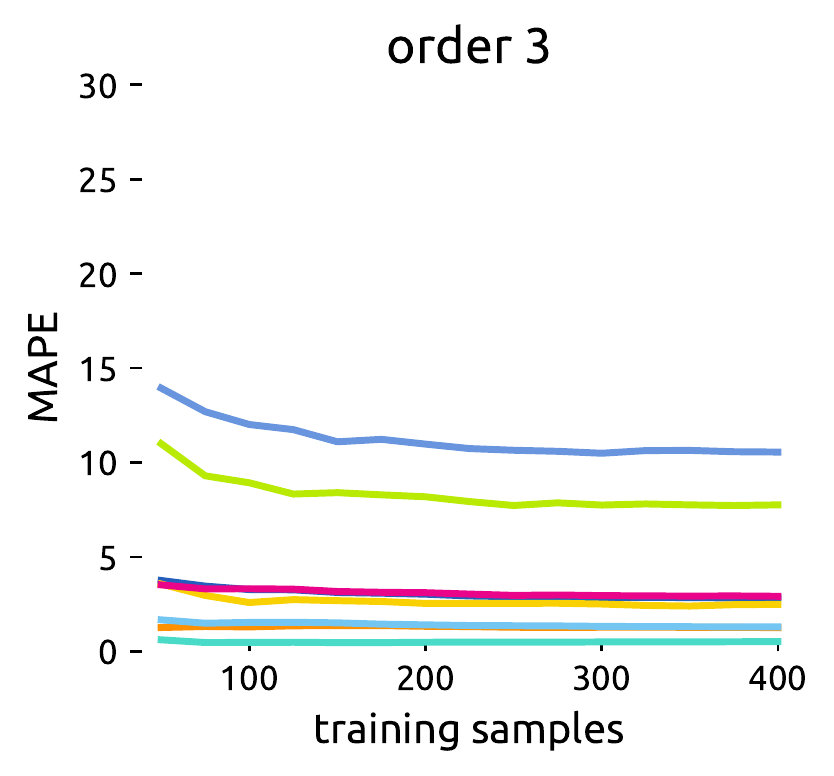}
            \includegraphics[height=.2\textwidth, trim=0cm 0cm 4.5cm 0cm, clip]{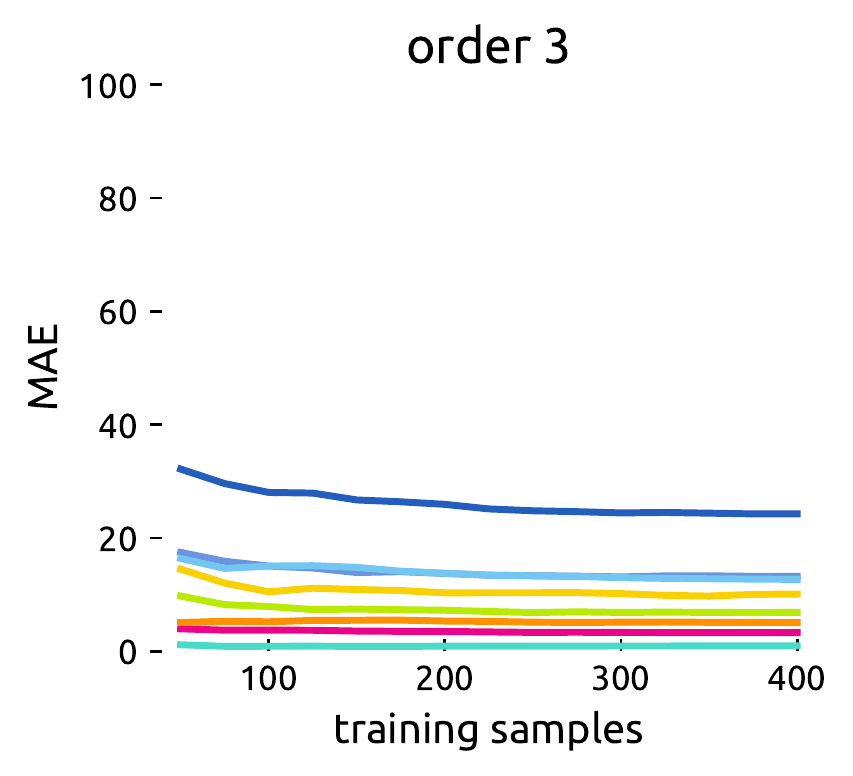}
            \includegraphics[height=.2\textwidth]{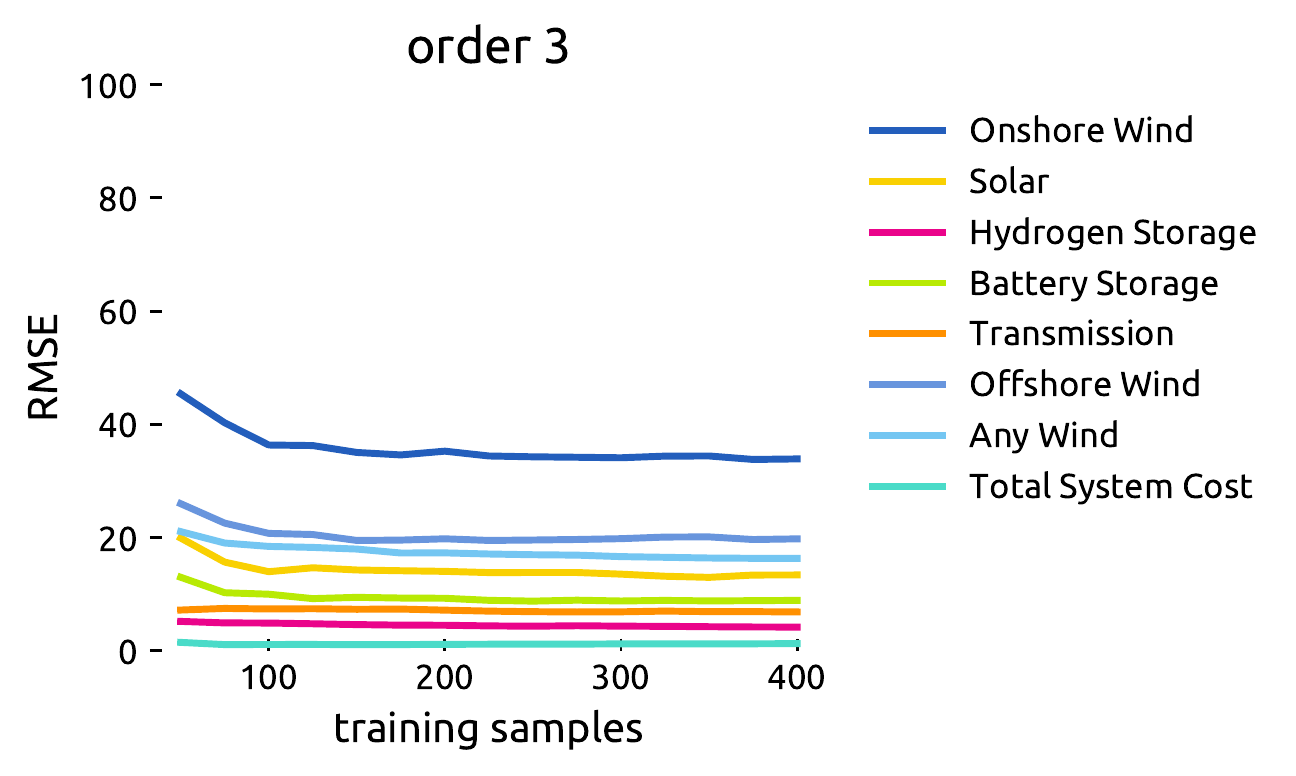}
        \end{subfigure}
    }
    \noindent\makebox[\textwidth]{
        \begin{subfigure}[t]{1.4\textwidth}
            \caption{polynomial order}
            \label{fig:error:poly}
            \includegraphics[height=.2\textwidth, trim=0cm 0cm 4.5cm 0cm, clip]{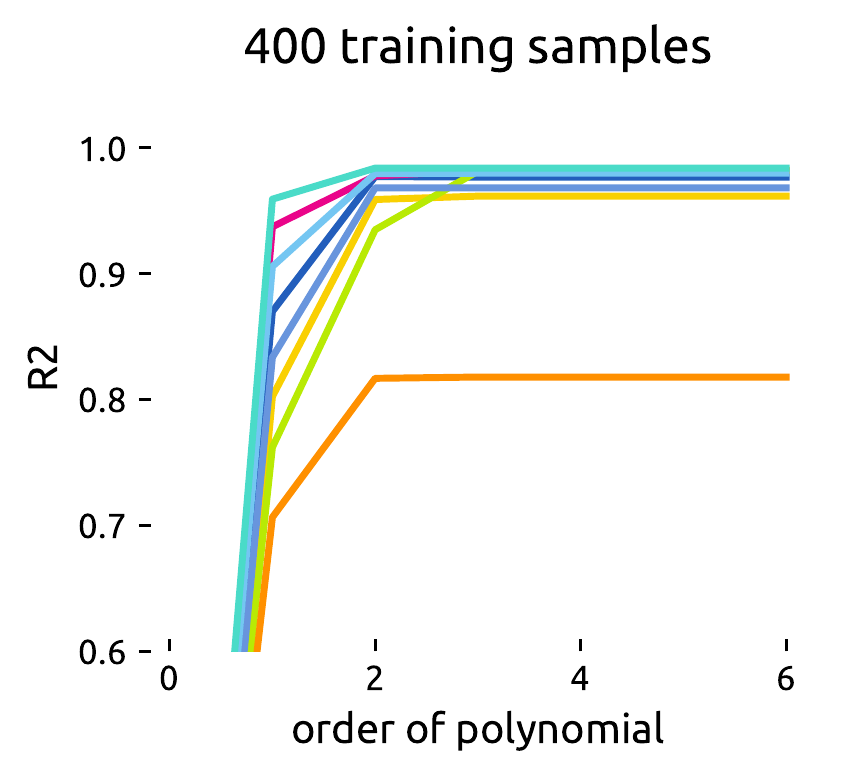}
            \includegraphics[height=.2\textwidth, trim=0cm 0cm 4.5cm 0cm, clip]{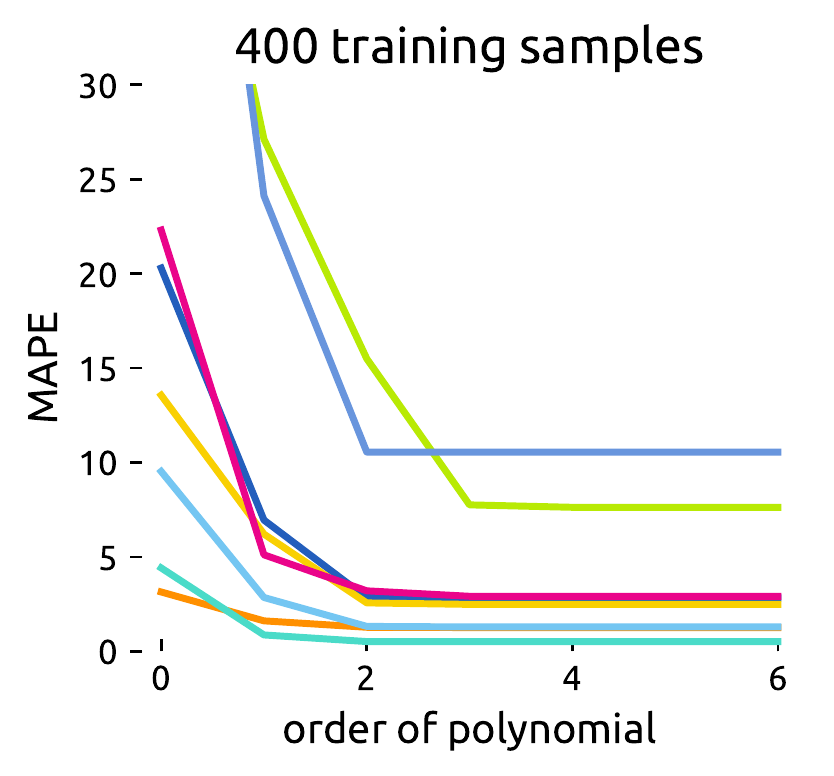}
            \includegraphics[height=.2\textwidth, trim=0cm 0cm 4.5cm 0cm, clip]{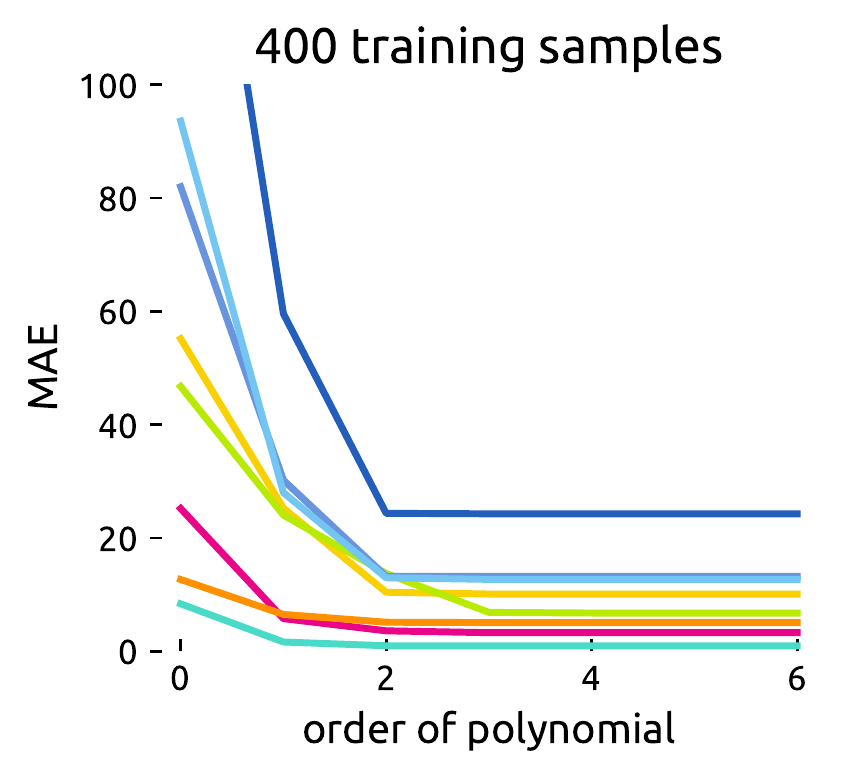}
            \includegraphics[height=.2\textwidth]{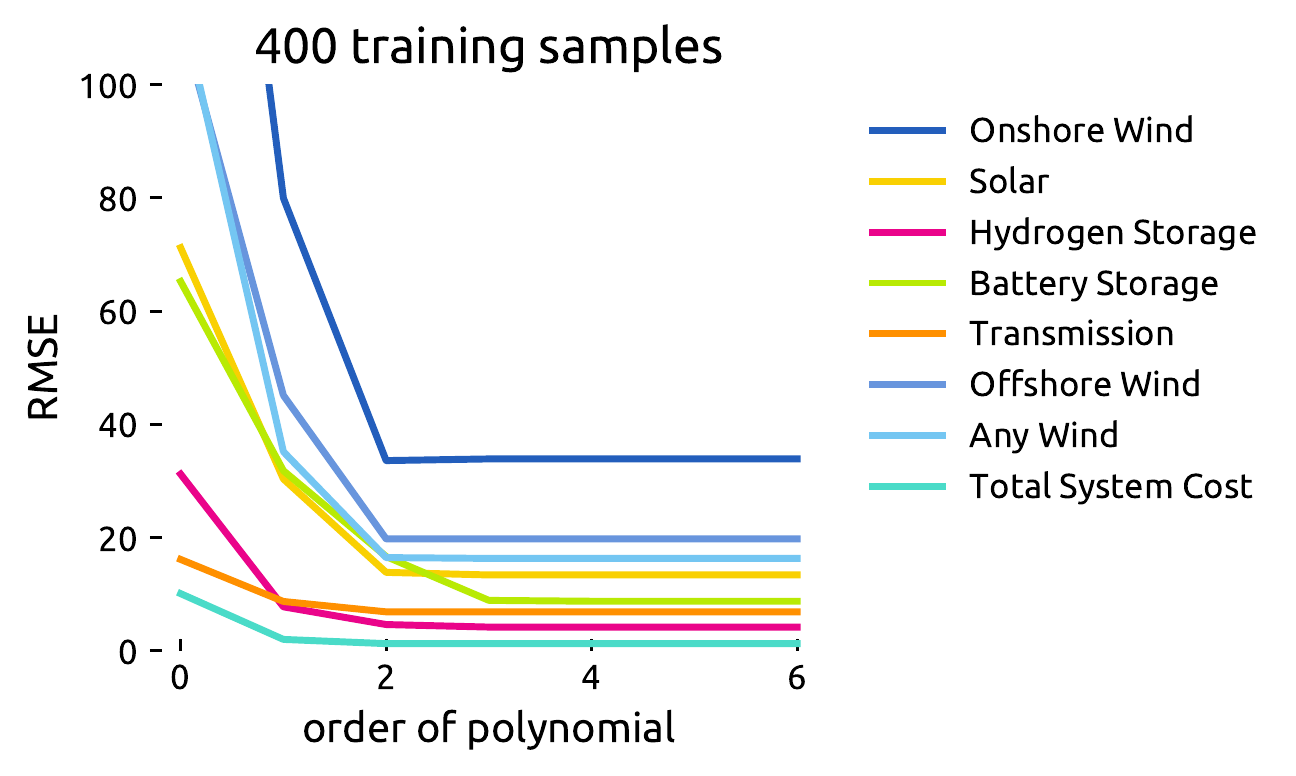}
        \end{subfigure}
    }
    \caption[Cross-validation errors]{Cross-validation errors by output for varying sample sizes and polynomial orders
    of least-cost low-fidelity surrogate models.}
    \label{fig:error}
\end{figure}

We justify the use of surrogate modelling by cross-validation.
Out of the 500 low-fidelity samples,
100 samples are not used in the regression.
This validation dataset is unknown to the surrogate model and is
consulted to assess the approximation's quality.
Because the high-fidelity sample size is limited and
approximating near-optimal solutions is not assumed to fundamentally differ,
we base the validation on low-fidelity least-cost solutions only.
We experimentally evaluate the approximation errors between predicted and observed data
for different combinations of polynomial order and sample size
to decide on a suitable parameterisation.
We present the coefficient of determination (R$^2$)
for the variance captured,
the mean absolute (percentage) errors (MAE/MAPE) for absolute and relative deviations,
and the root mean squared error (RMSE).

Regarding the number of samples required, \cref{fig:error:samples}
foremost illustrates that, given enough samples, we achieve
average relative errors of less than 4\% for most output variables.
This is comparable to the cross-validation errors from Tröndle et al.~\cite{trondle_trade-offs_2020}
at rates below 5\%.
Only for offshore wind and battery storage, we observe larger errors.
However, this can be explained by a distortion of the relative measure
when these technologies are hardly built for some cost projections.
On the contrary, the prediction of total system costs is remarkably accurate.
\cref{fig:error:samples} also demonstrates that for a polynomial order of 3,
we gain no significant improvement with more than 200 samples.
In fact, thanks to the regularisation term used in the regression,
we already attain acceptable levels of accuracy with as few as 50 samples.
Moreover, the high R$^2$ values underline that the surrogate model can explain
most of the output variance.

Regarding the polynomial order, \cref{fig:error:poly} shows that
an order of 2 and below may be too simple to capture the
interaction between different parameters. On the other hand,
an order of 4 and above yields no improvement and,
were it not for the moderating regularisation term,
would even result in a loss of generalisation properties due to overfitting.
As higher-order approximations require significantly more samples,
an order of 3 appears to be a suitable compromise
to limit the computational burden.

\section{Results and Discussion}
\label{sec:results}

In this section, we approach the uncertainty analysis to near-optimal solutions
by reviewing the propagation of input uncertainties into least-cost solutions
first and expanding gradually from there. This includes inspecting cost and
capacity distributions induced by unknown future technology cost and conducting
a global sensitivity analysis that identifies the most influential cost
parameters for least-cost solutions. We then expand the uncertainty analysis to
the space of nearly cost-optimal solutions, which yields us insights about the
consistency of near-optimal alternatives across a variety of cost parameters.

\subsection{Cost and Capacity Distribution of Least-Cost Solutions}

\begin{SCfigure}
    \includegraphics[width=0.75\textwidth]{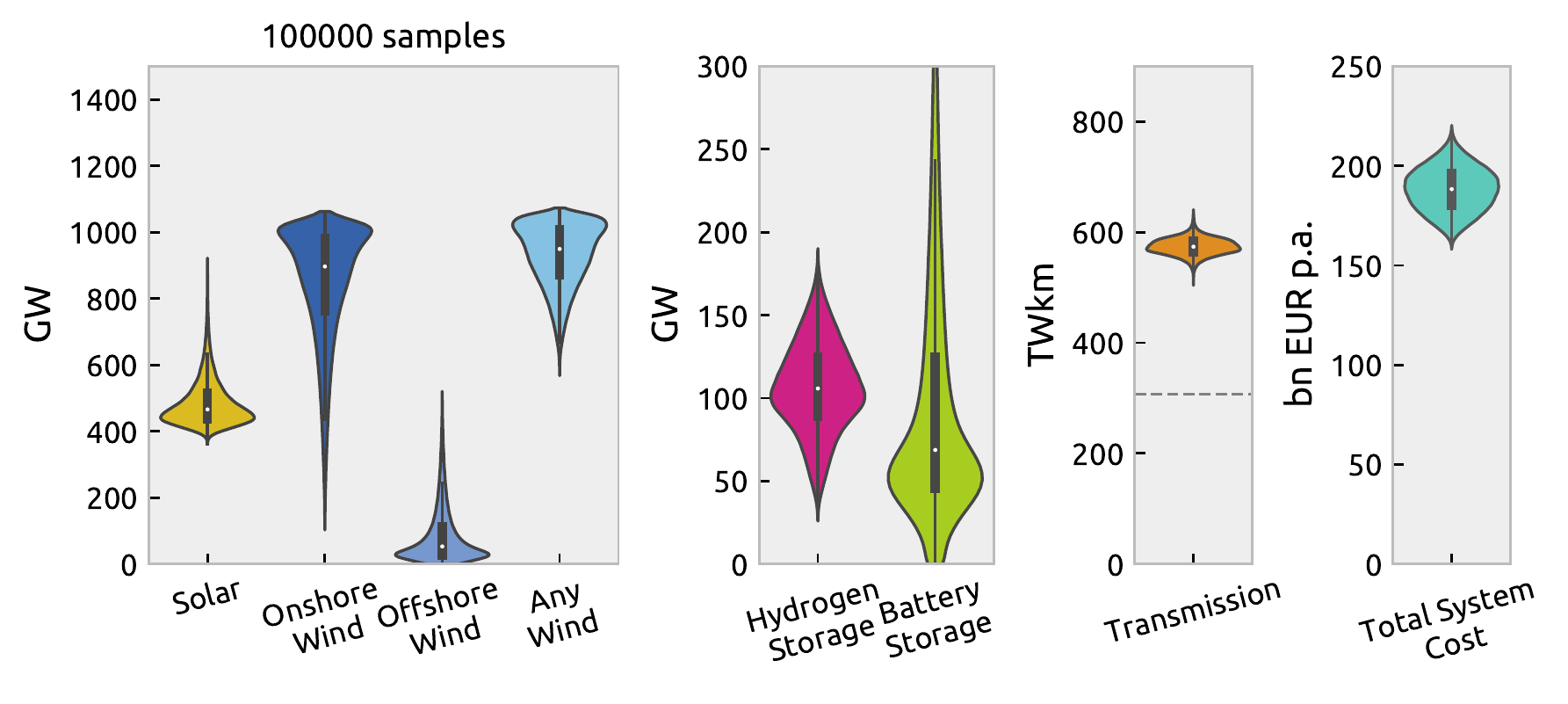}
    \caption{
      Distribution of total system cost, generation, storage, and transmission capacities
      for least-cost solutions.
    }
    \label{fig:violin}
\end{SCfigure}

Based on the uncertainty of cost inputs, the total annual system costs vary
between 160 and 220 billion Euro per year, as displayed in \cref{fig:violin}.
This means the most pessimistic cost projections entail about 40\% higher cost
than the most optimistic projections. All least-cost solutions build at least
350 GW solar and 600 GW wind, but no more than 1100 GW. While wind capacities
tend towards higher values, solar capacities tend towards lower values. We
observe that least-cost solutions clearly prefer onshore over offshore wind, yet
onshore wind features the highest uncertainty range alongside battery storage.
The cost optimum gravitates towards hydrogen storage rather than battery storage
unless battery storage becomes very cheap. There are no least-cost solutions
without hydrogen, only some without battery storage. Transmission network
expansion is least affected by cost uncertainty and consistently doubled
compared to today's capacities. The question arises, what we can conclude from
these insights. The interpretation of the observed ranges may be limited because
they are not robust when we look beyond the least-cost solutions and acknowledge
structural modelling uncertainties, such as social constraints. Moreover, the
pure distribution of outputs does not yet convey information about how sensitive
results are to particular cost assumptions. But knowing the technologies for
which lowering overnight costs has a significant impact is important to promote
technological learning in that direction.

\begin{figure}
    \begin{subfigure}[t]{0.32\textwidth}
        \caption{onshore wind}
        \includegraphics[width=\textwidth]{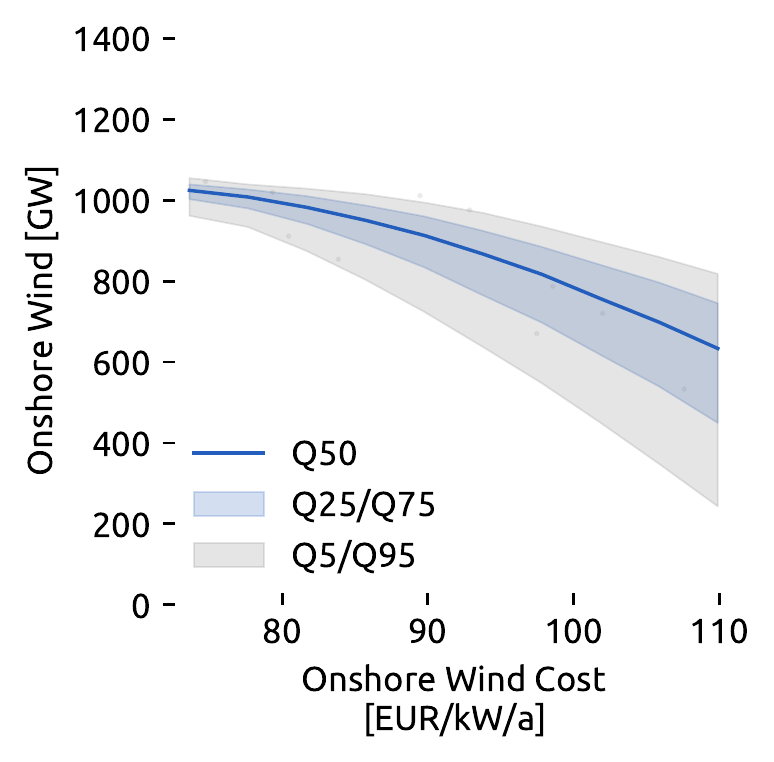}
    \end{subfigure}
    \begin{subfigure}[t]{0.32\textwidth}
        \caption{offshore wind}
        \includegraphics[width=\textwidth]{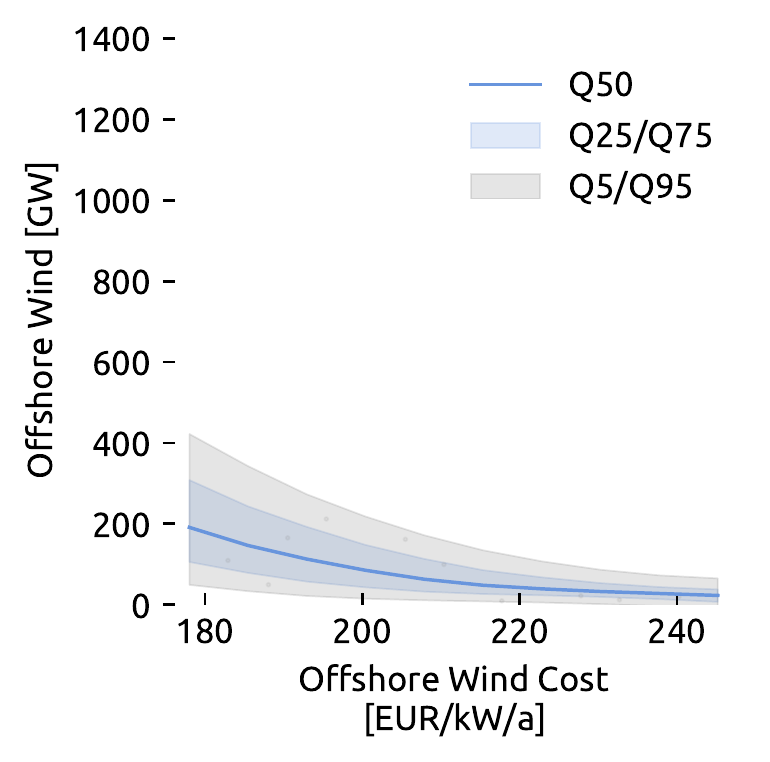}
    \end{subfigure}
    \begin{subfigure}[t]{0.32\textwidth}
        \caption{solar}
        \includegraphics[width=\textwidth]{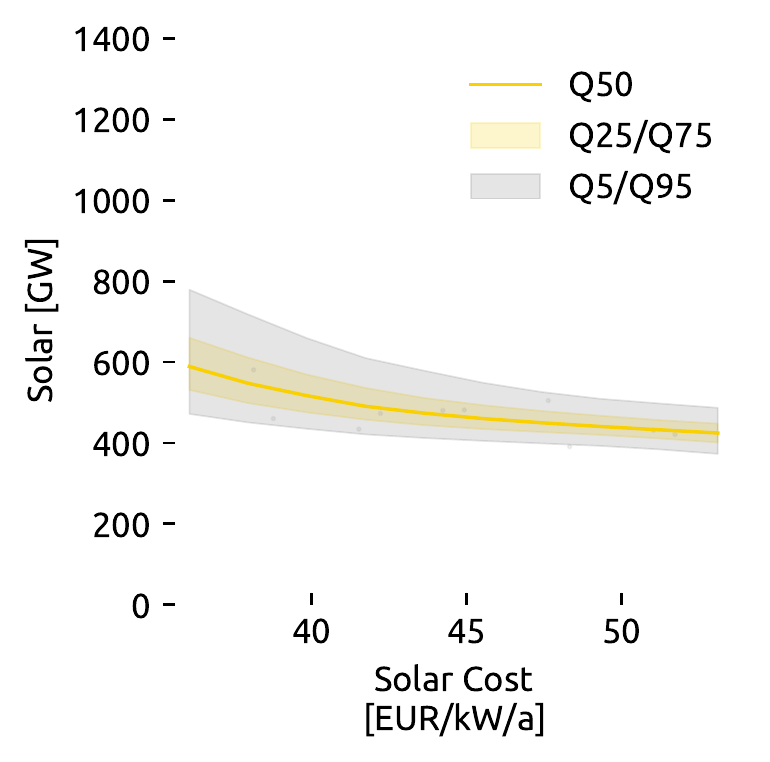}
    \end{subfigure} \\
    \begin{subfigure}[t]{0.32\textwidth}
        \caption{battery storage}
        \includegraphics[width=\textwidth]{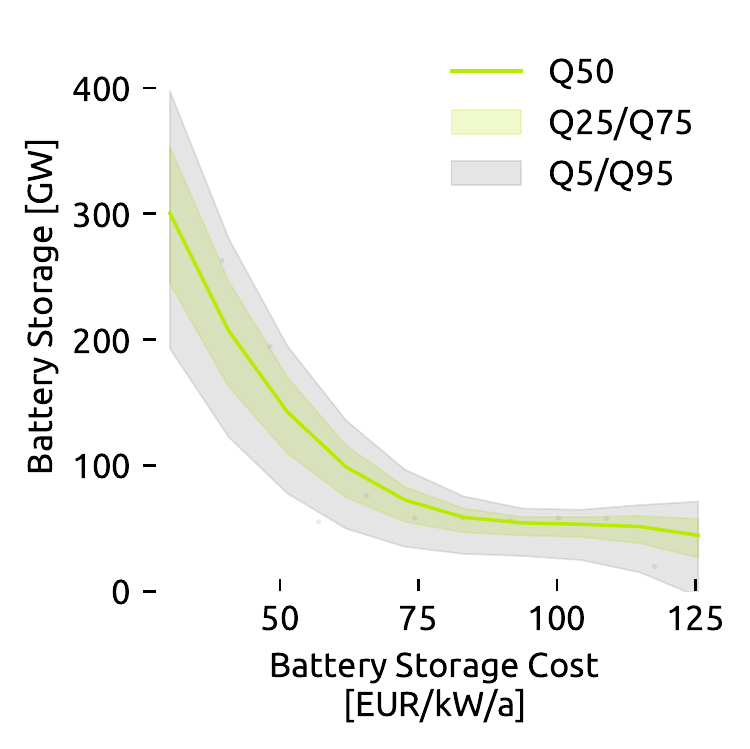}
    \end{subfigure}
    \begin{subfigure}[t]{0.32\textwidth}
        \caption{hydrogen storage}
        \includegraphics[width=\textwidth]{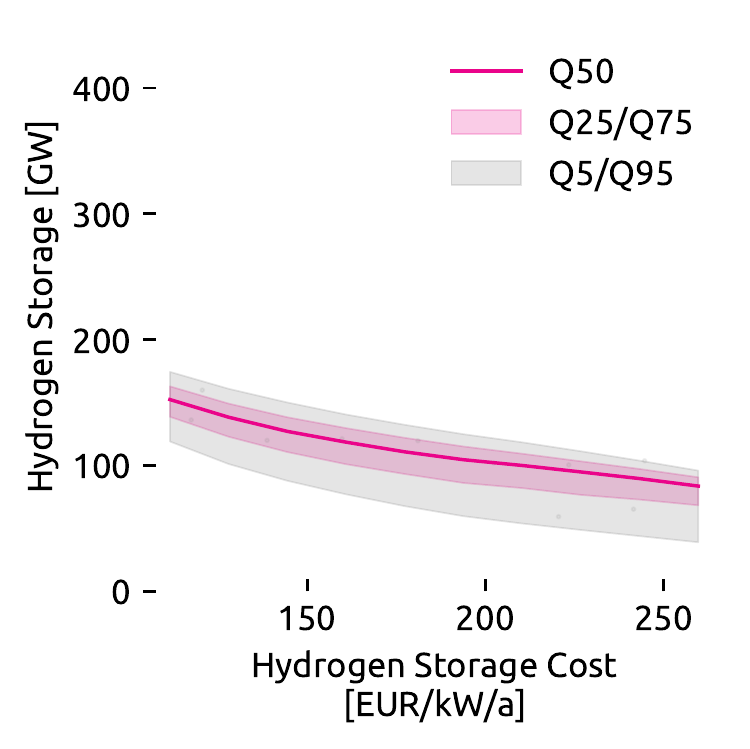}
    \end{subfigure}
    \begin{subfigure}[t]{0.32\textwidth}
        \caption{transmission}
        \includegraphics[width=\textwidth]{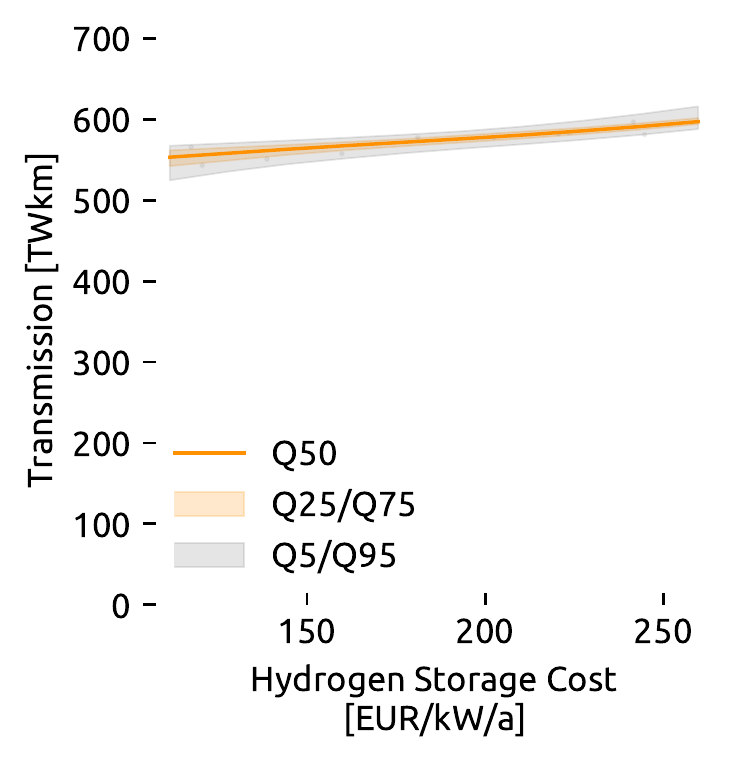}
    \end{subfigure}
    \vspace{-0.3cm}
    \caption{
      Sensitivity of capacities towards their own technology cost.
      The median (Q50) alongside the 5\%, 25\%, 75\%, and 95\% quantiles (Q5--Q95) display
      the sensitivity subject to the uncertainty induced by other cost parameters.
    }
    \label{fig:sensitivity}
\end{figure}

\begin{figure}
    \begin{subfigure}[t]{0.45\textwidth}
        \caption{first-order Sobol indices [\%]}
        \label{fig:sobol:first}
        \includegraphics[width=\textwidth]{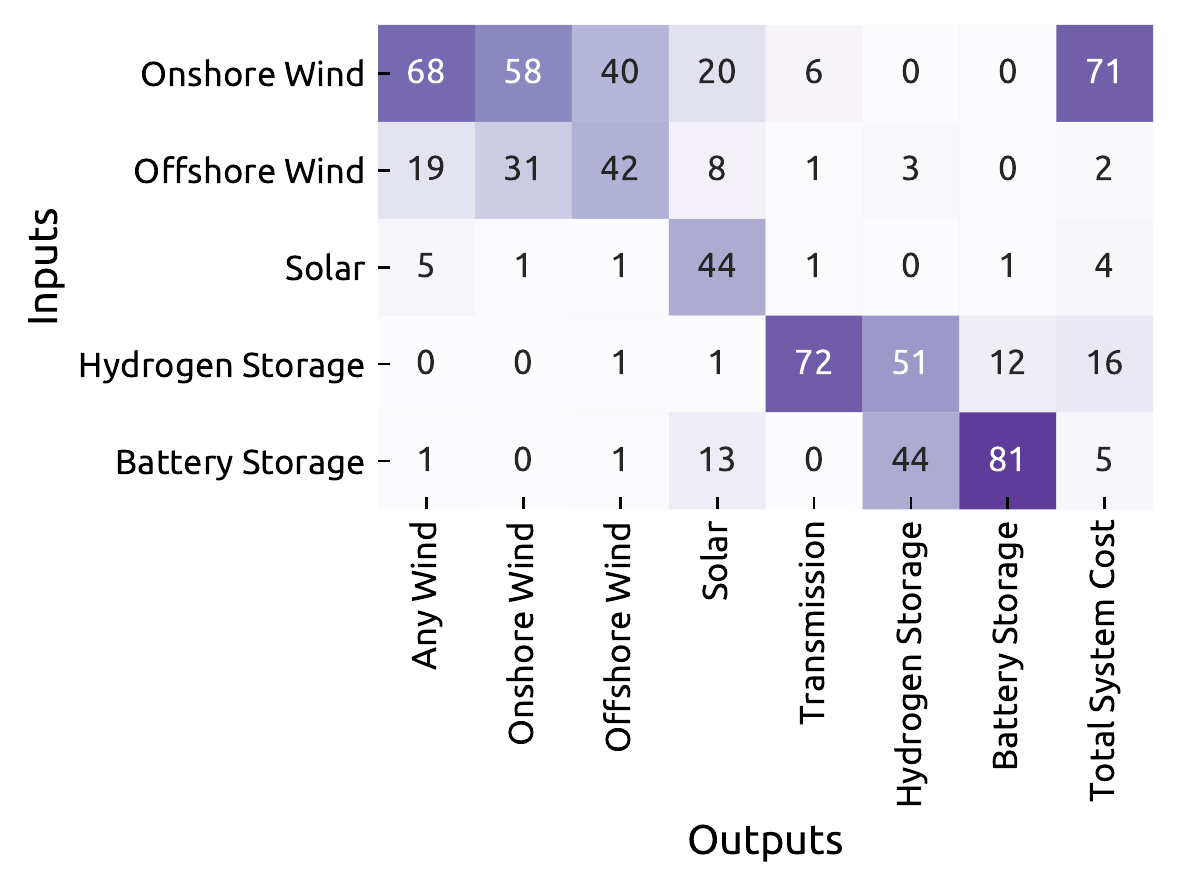}
    \end{subfigure}
    \begin{subfigure}[t]{0.54\textwidth}
        \caption{total Sobol indices [\%]}
        \label{fig:sobol:total}
        \includegraphics[width=\textwidth]{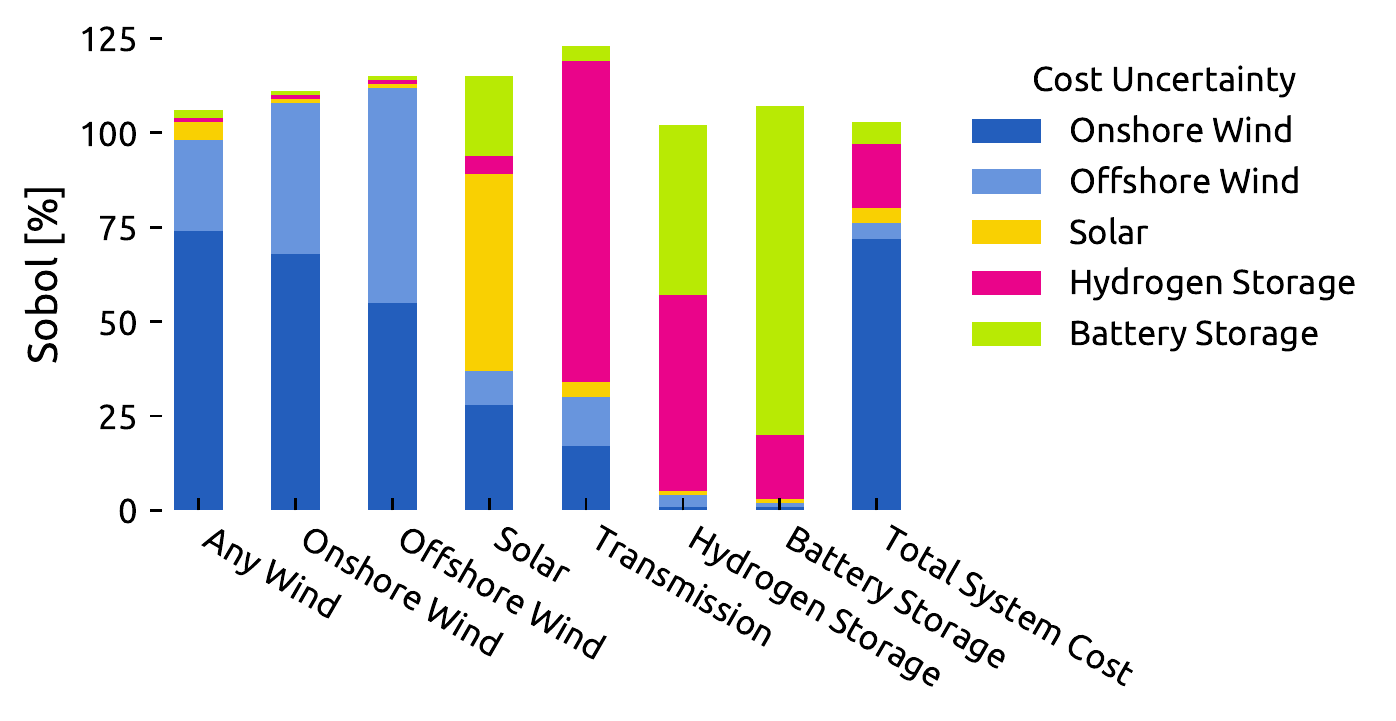}
    \end{subfigure}
    \vspace{-0.3cm}
    \caption[First-order and total Sobol indices]{
      Sobol indices. These sensitivity indices attribute output variance to random input variables
      and reveal which inputs the outputs are most sensitive to. The first-order Sobol indices
      quantify the share of output variance due to variations in one input parameter alone.
      The total Sobol indices further include interactions with other input variables.
      Total Sobol indices can be greater than 100\% if the contributions are not purely additive.
    }
    \label{fig:sobol}
\end{figure}

\subsection{Parameter Sweeps and Global Sensitivity Indices}

\cref{fig:sensitivity} addresses a selection of local self-sensitivities,
i.e.~how the cost of a technology influences its deployment while displaying the
remaining uncertainty induced by other cost parameters. The overall tendency is
easily explained: the cheaper a technology becomes, the more it is built.
However, changes of slope and effects on the uncertainty range as one cost
parameter is swept are insightful nonetheless. For instance,
\cref{fig:sensitivity} reveals that battery storage becomes significantly more
attractive economically once its annuity falls below 75 EUR/kW/a (including 6h
energy capacity at full power output) hydrogen storage features a steady slope.
A low cost of onshore wind makes building much onshore wind capacity attractive
with low uncertainty, whereas if onshore wind costs are high how much is built
greatly depends on other cost parameters.
The opposite behaviour is observed for offshore wind and solar.
The cost of hydrogen storage mostly causes the limited uncertainty about cost-optimal levels of grid expansion.
As the cost of hydrogen storage falls, less grid reinforcement is chosen.
But since the presented self-sensitivities only exhibit a fraction of all sensitivities,
in the next step we formalise how input uncertainties affect each outcome
systematically by applying variance-based global sensitivity analysis techniques, which
have been applied in the context of energy systems e.g.~in \cite{trondle_trade-offs_2020,mavromatidis_uncertainty_2018}.

Sensitivity indices, or Sobol indices, attribute the observed output variance to each input and
can be computed analytically from the polynomial chaos expansion \cite{sudret_global_2008}.
For our application, the Sobol indices can, for instance, tell us which technology cost
contributes the most to total system cost or how much of a specific technology will be built.
The first-order Sobol indices describe the share of output variance
due to variations in one input alone averaged over variations in the other inputs.
Total Sobol indices also consider higher-order interactions,
which are greater than 100\% if the relations are not purely additive.

The first-order and total Sobol indices for least-cost solutions in
\cref{fig:sobol} show that the total system cost is largely determined by how
expensive it is to build onshore wind capacity, followed by the cost of hydrogen
storage. The amount of wind in the system is almost exclusively governed by the
cost of onshore and offshore wind parks. Other carriers yield a more varied
picture. The cost-optimal solar capacities additionally depend on onshore wind
and battery costs. The amount of hydrogen storage is influenced by battery and
hydrogen storage cost alike.  Although there are noticeable higher-order
effects, which are most extensive for transmission, the first-order effects
dominate. Strikingly, the volume of transmission network expansion strongly
depends on the cost of hydrogen storage, which can be explained by the synoptic
spatio-temporal scale of wind power variability across the European continent
which both hydrogen storage and transmission networks seek to balance from
different angles. While hydrogen storage typically balances multi-week
variations in time, continent-spanning transmission networks exploit the
circumstance that as weather systems traverse the continent, it is likely always
to be windy somewhere in Europe.

\subsection{Fuzzy Near-Optimal Corridors with Increasing Cost Slack}

So far, we quantified the output uncertainty and analysed the sensitivity
towards inputs at least-cost solutions only. Yet, it has been previously shown
that even for a single cost parameter set a wide array of technologically
diverse but similarly costly solutions exists \cite{nearoptimal}. In the next
step, we examine how technology cost uncertainty affects the shape of the space
of near-optimal solutions.

\begin{figure}
    \vspace{-2cm}
    \noindent\makebox[\textwidth]{
    \begin{subfigure}[t]{0.45\textwidth}
        \centering
        \includegraphics[width=\textwidth]{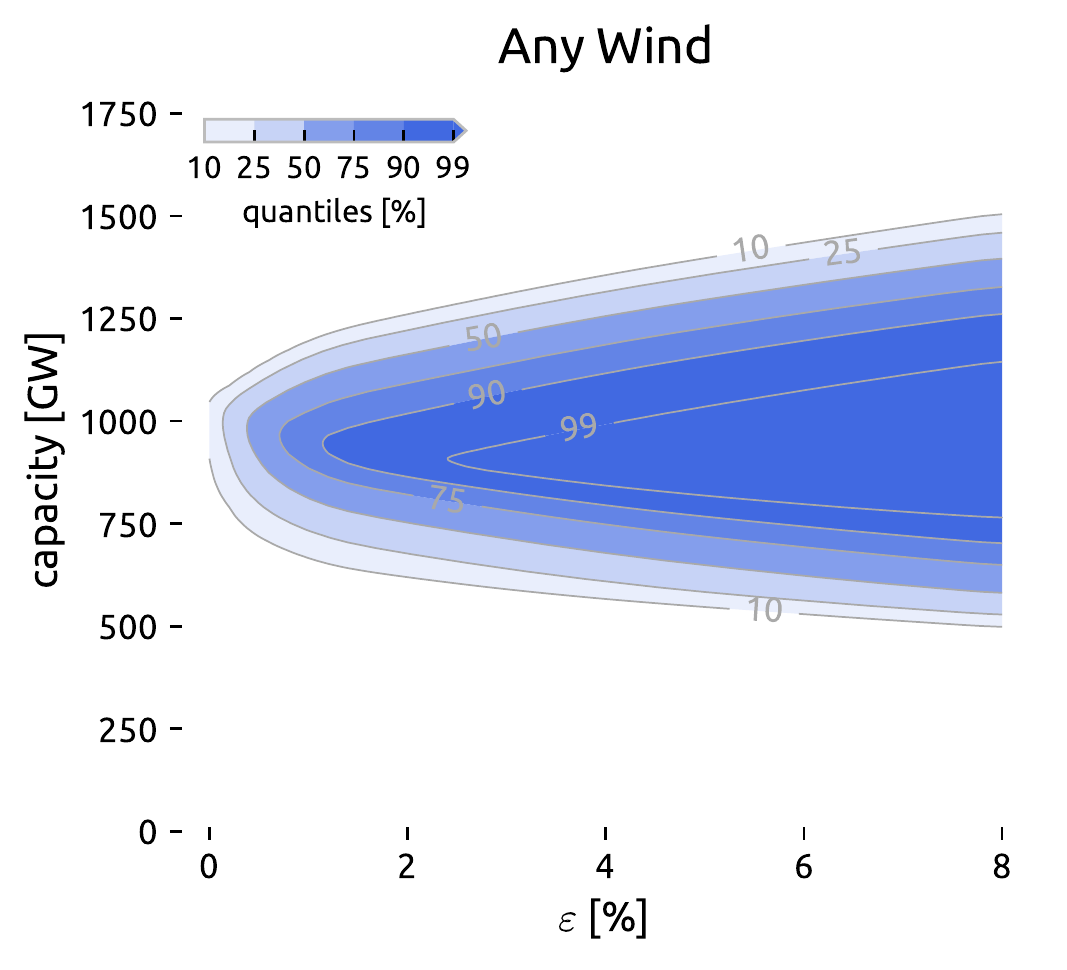}
    \end{subfigure}
    \begin{subfigure}[t]{0.45\textwidth}
        \centering
        \includegraphics[width=\textwidth]{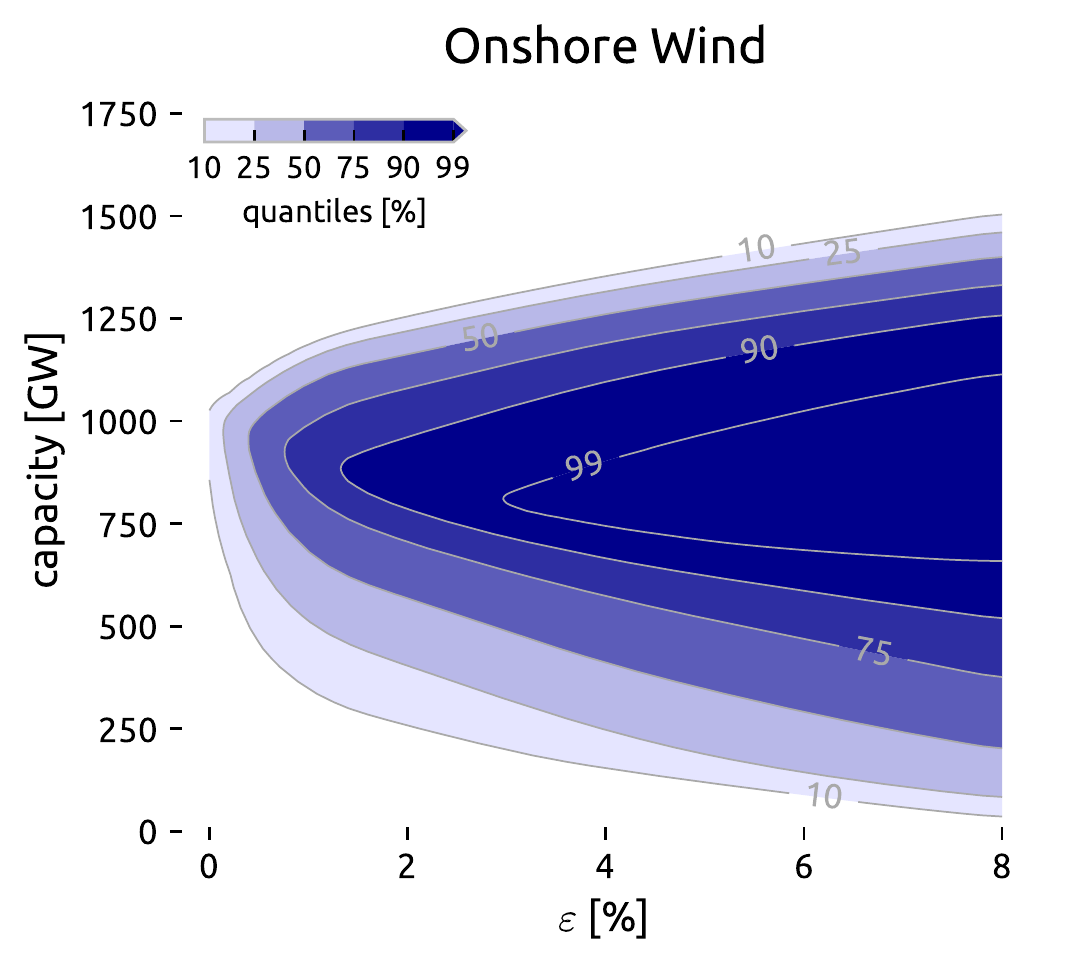}
    \end{subfigure}
    \begin{subfigure}[t]{0.45\textwidth}
        \centering
        \includegraphics[width=\textwidth]{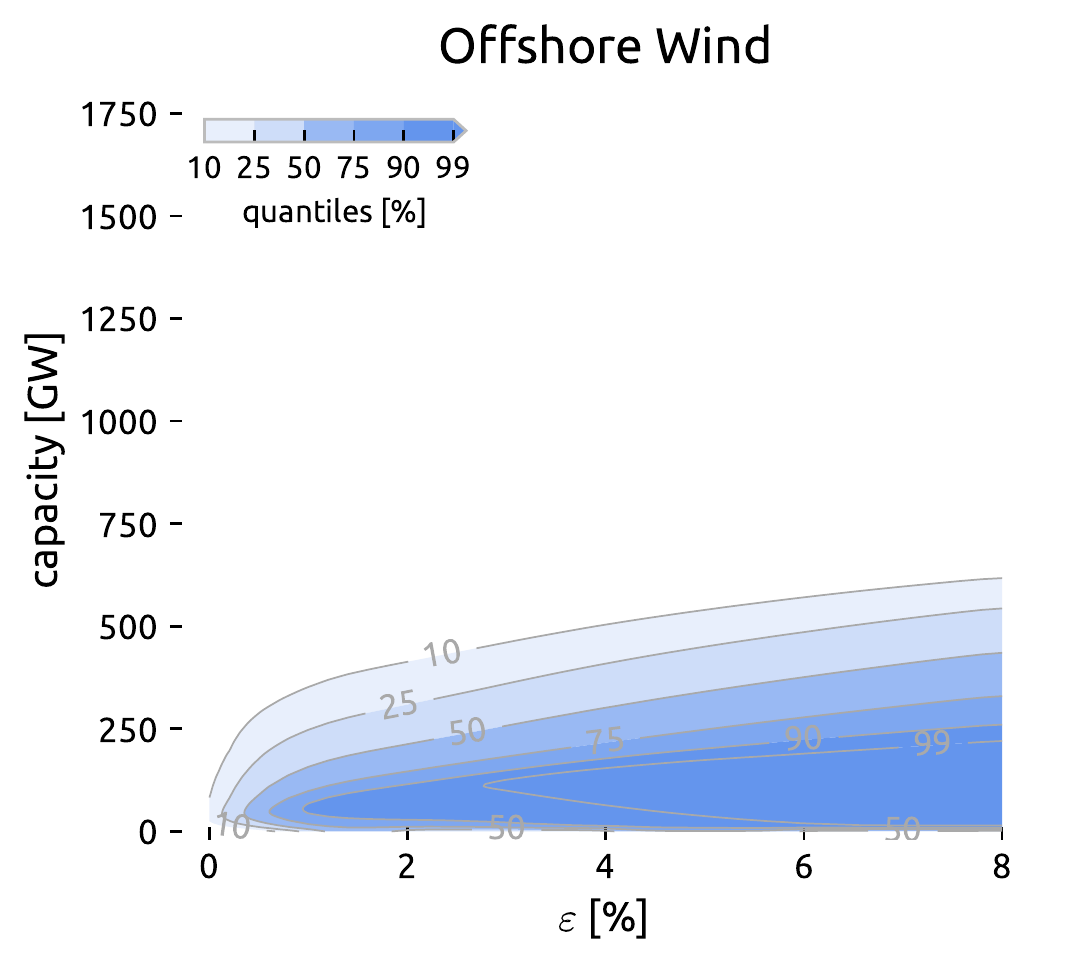}
    \end{subfigure}
    }
    \noindent\makebox[\textwidth]{
    \begin{subfigure}[t]{0.45\textwidth}
        \centering
        \includegraphics[width=\textwidth]{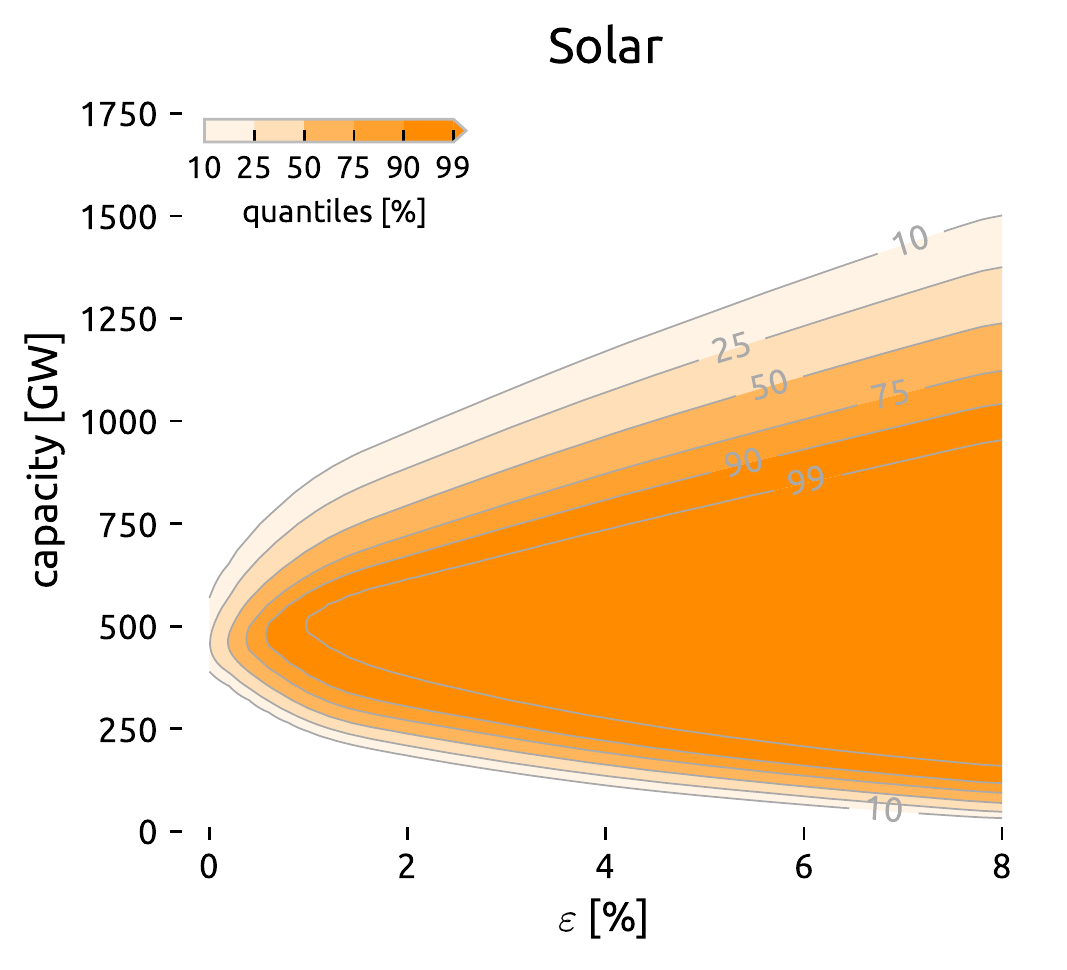}
    \end{subfigure}
    \begin{subfigure}[t]{0.45\textwidth}
        \centering
        \includegraphics[width=\textwidth]{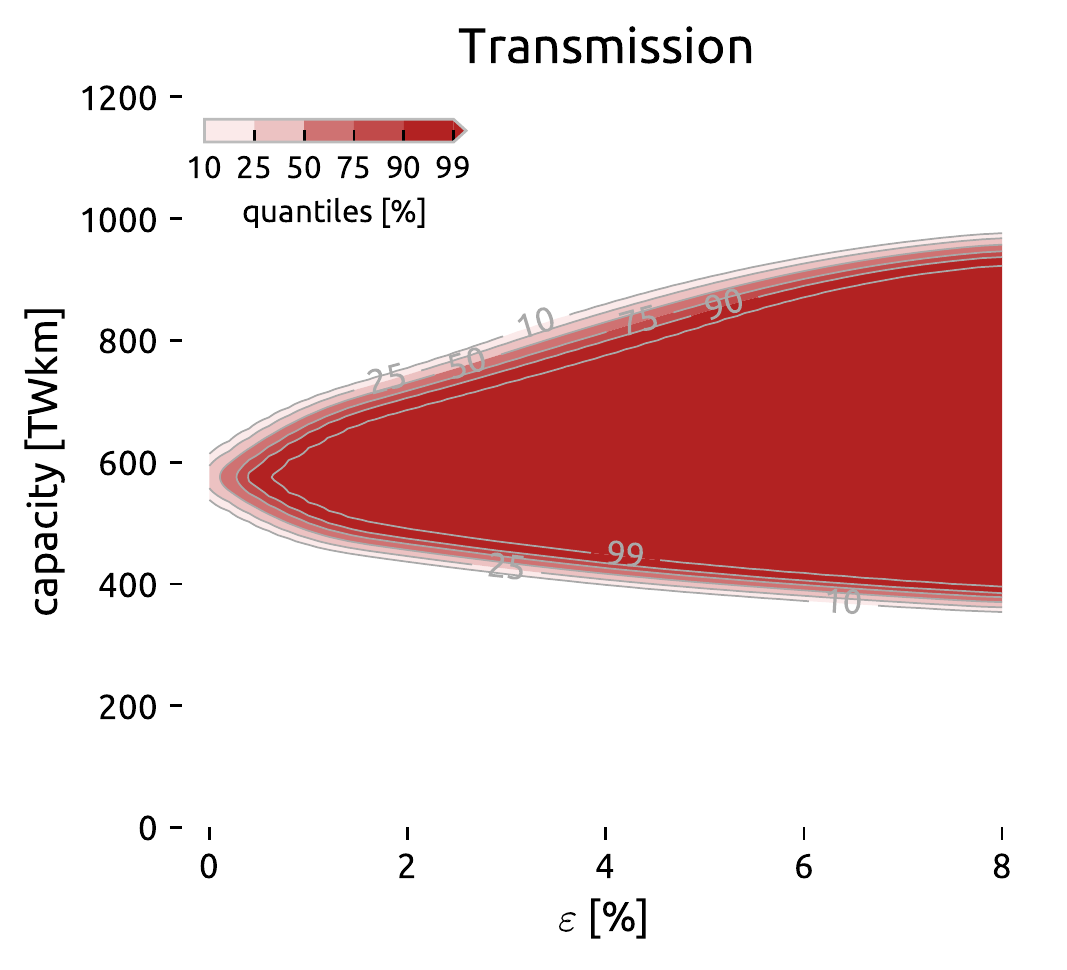}
    \end{subfigure}
    }
    \noindent\makebox[\textwidth]{
        \begin{subfigure}[t]{0.45\textwidth}
            \centering
            \includegraphics[width=\textwidth]{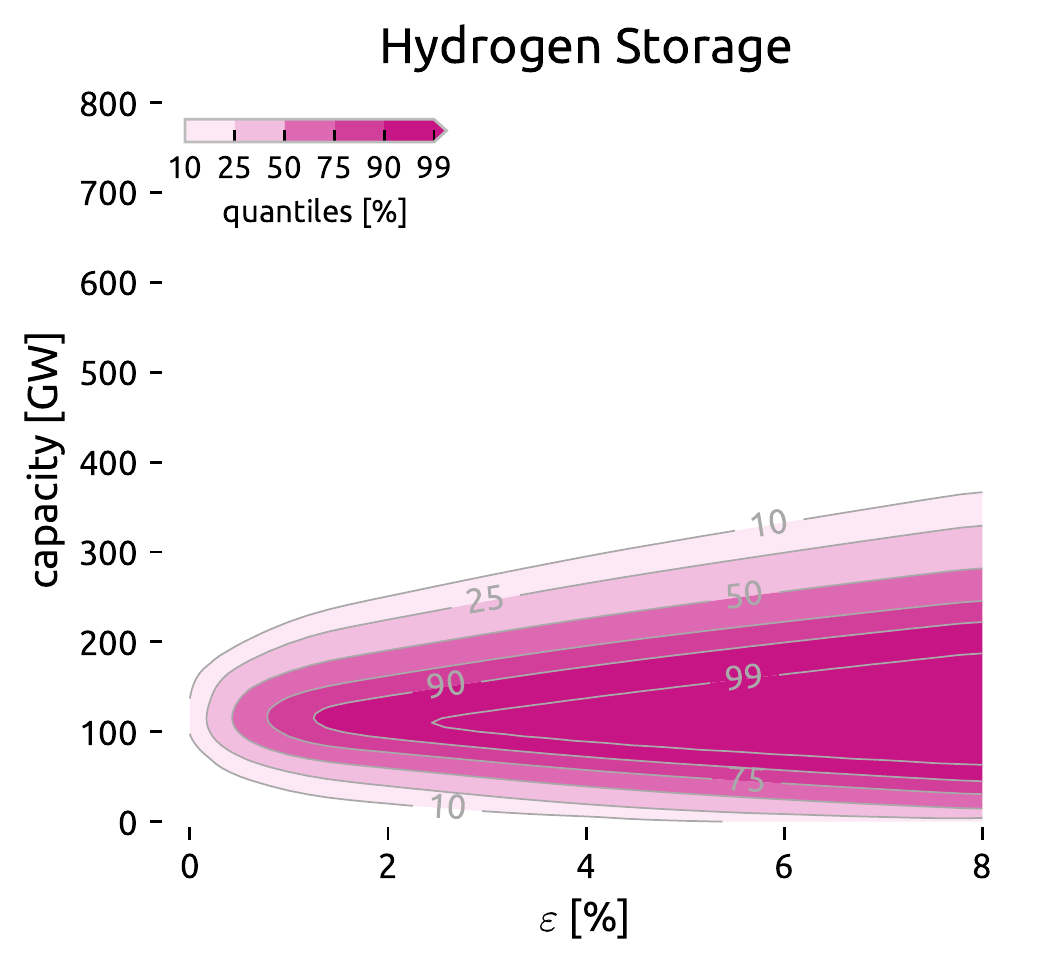}
        \end{subfigure}
        \begin{subfigure}[t]{0.45\textwidth}
            \centering
        \includegraphics[width=\textwidth]{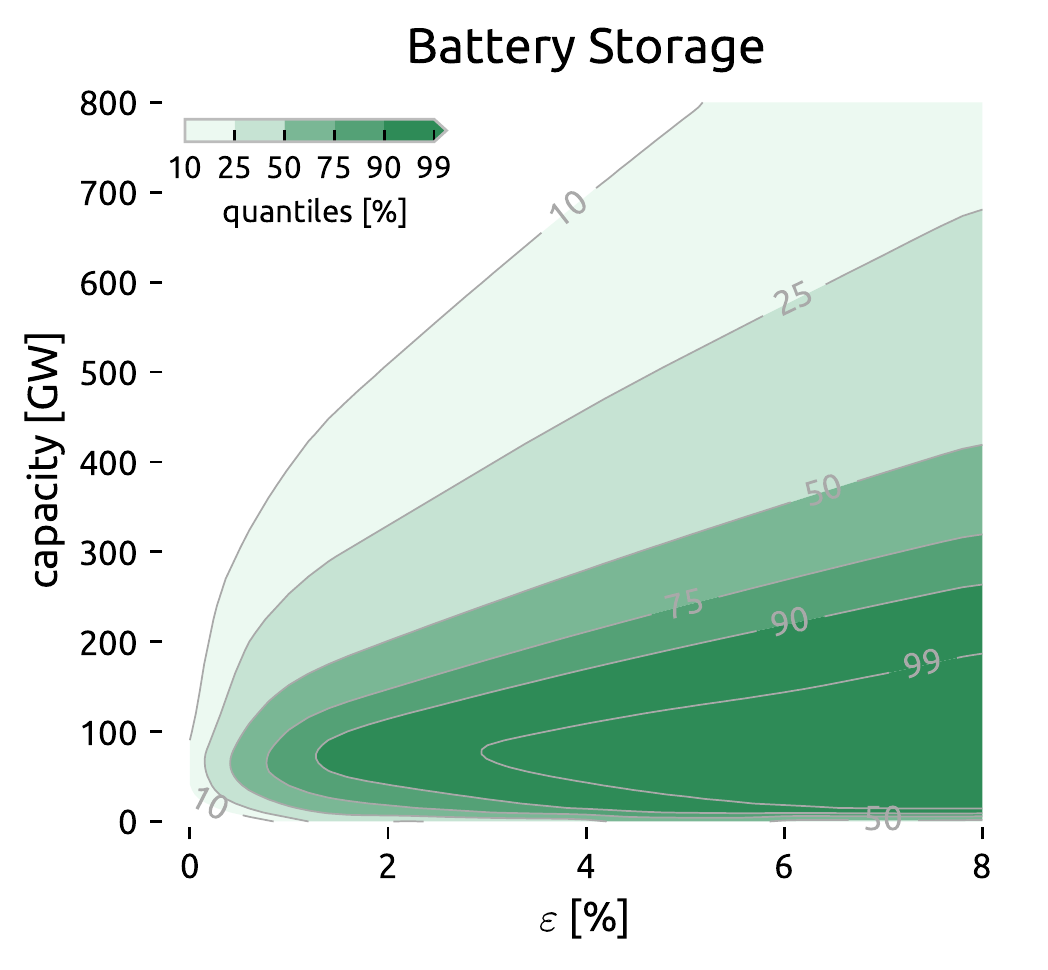}
    \end{subfigure}
    }
    \caption{
    Space of near-optimal solutions by technology under cost uncertainty.
    For each technology and cost sample,
    the minimum and maximum capacities obtained for increasing cost penalties
    $\epsilon$ form a cone, starting from a common least-cost solution.
    By arguments of convexity, the capacity ranges contained by the cone can be near-optimal and feasible, given a degree of freedom in the other technologies.
    From optimisation theory, we know that the cones widen up for increased slacks.
    As we consider technology cost uncertainty, the cone will look slightly different for each sample.
    The contour lines represent the frequency a solution is inside the near-optimal cone over the whole parameter space.
    This is calculated from the overlap of many cones, each representing a set of cost assumptions.
    Due to discrete sampling points in the $\epsilon$-dimension, the plots further apply quadratic interpolation and a Gaussian filter for smoothing.
    }
    \label{fig:fuzzycone}
\end{figure}

By identifying feasible alternatives common to all, few or no cost samples, we
outline low-cost solutions common to most parameter sets (e.g. above 90\%
contour) as well as system layouts that do not meet low-cost criteria in any
circumstances for varying $\epsilon$ in \cref{fig:fuzzycone}. The wider the
displayed contour lines are apart, the more uncertainty exists about the
boundaries. The closer contour lines are together, the more specific the limits
are. The height of the quantiles quantifies flexibility for a given level of
certainty and slack; the angle presents information about the sensitivity
towards cost slack.

From the fuzzy upper and lower Pareto fronts in \cref{fig:fuzzycone} we can see that it is
highly likely that building 900 GW of wind capacity is possible within 3\% of the optimum, and that
conversely building less than 600 GW has a low chance of being near the cost optimum.
Only a few solutions can forego onshore wind entirely and remain within 8\% of the cost-optimum,
whereas it is very likely possible to build a system without offshore wind at a cost penalty of 4\% at most.
On the other hand, more offshore wind generation is equally possible.
Unlike for onshore wind, where it is more uncertain how little can be built,
uncertainty regarding offshore wind deployment exists about how much can be built
so that costs remain within a pre-specified range.
For solar, the range of options within 8\% of the cost optimum at 90\% certainty is very wide.
Anything between 100 GW and 1000 GW appears feasible.
In comparison to onshore wind, the uncertainty about minimal solar requirements is smaller.

The level of required transmission expansion is least affected by the cost uncertainty.
To remain within $\epsilon=8\%$ it is just as likely possible to
plan for moderate grid reinforcement by 30\% as
is initiating extensive remodelling of the grid by tripling the transmission volume
compared to what is currently in operation.
These results indicate that in any case some transmission reinforcement
to balance renewable variations across the continent appears to be essential.
Hydrogen storage, symbolising long-term storage, also gives the impression of a vital technology in many cases.
Building 100 GW of hydrogen storage capacity is likely viable within 2\% of the cost optimum
and, even at $\epsilon=8\%$, only 25\% of cost samples require no long-term storage;
when battery costs are exceptionally low.
Overall, 90\% of cases appear to function without any short-term battery storage
while the system cost rises by 4\% at most.
However, especially battery storage exhibits a large degree of freedom to build more.

\subsection{Probabilistic Near-Optimal Space in Two Technology Dimensions}

\begin{figure}
    \noindent\makebox[\textwidth]{
    \begin{subfigure}[t]{0.45\textwidth}
        \centering
        \caption{wind and solar}
        \label{fig:dependencies:ws}
        \includegraphics[width=\textwidth]{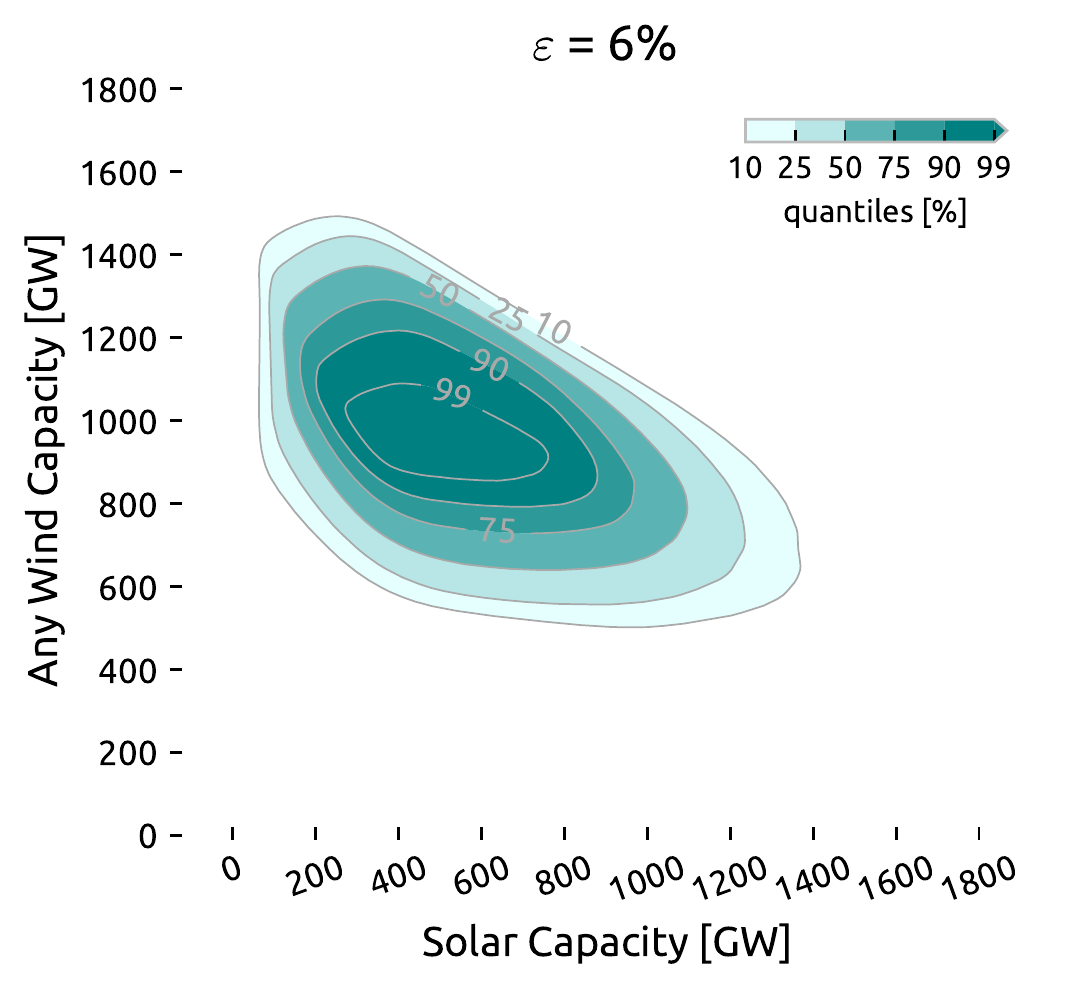}
    \end{subfigure}
    \begin{subfigure}[t]{0.45\textwidth}
        \centering
        \caption{offshore and onshore wind}
        \label{fig:dependencies:oo}
        \includegraphics[width=\textwidth]{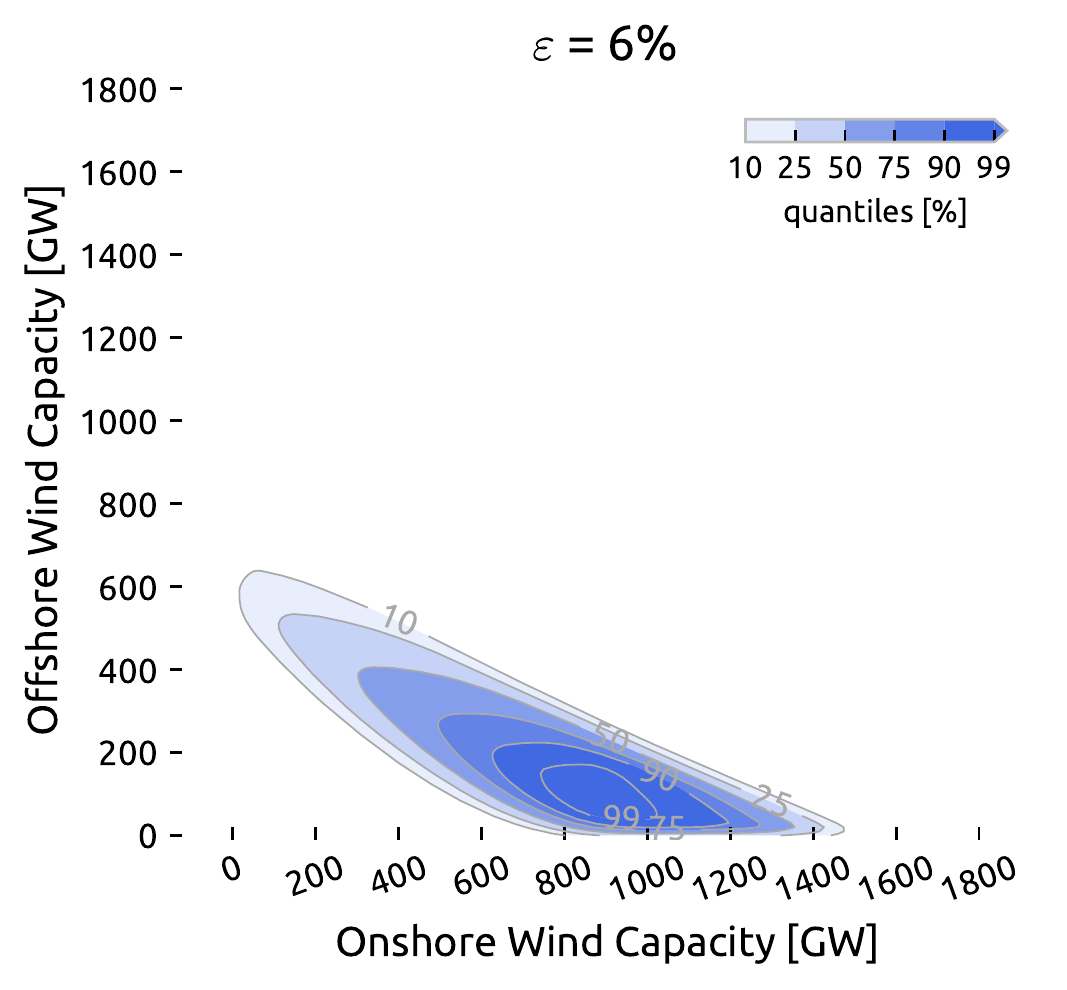}
    \end{subfigure}
    \begin{subfigure}[t]{0.45\textwidth}
        \centering
        \caption{hydrogen and battery storage}
        \label{fig:dependencies:hb}
        \includegraphics[width=\textwidth]{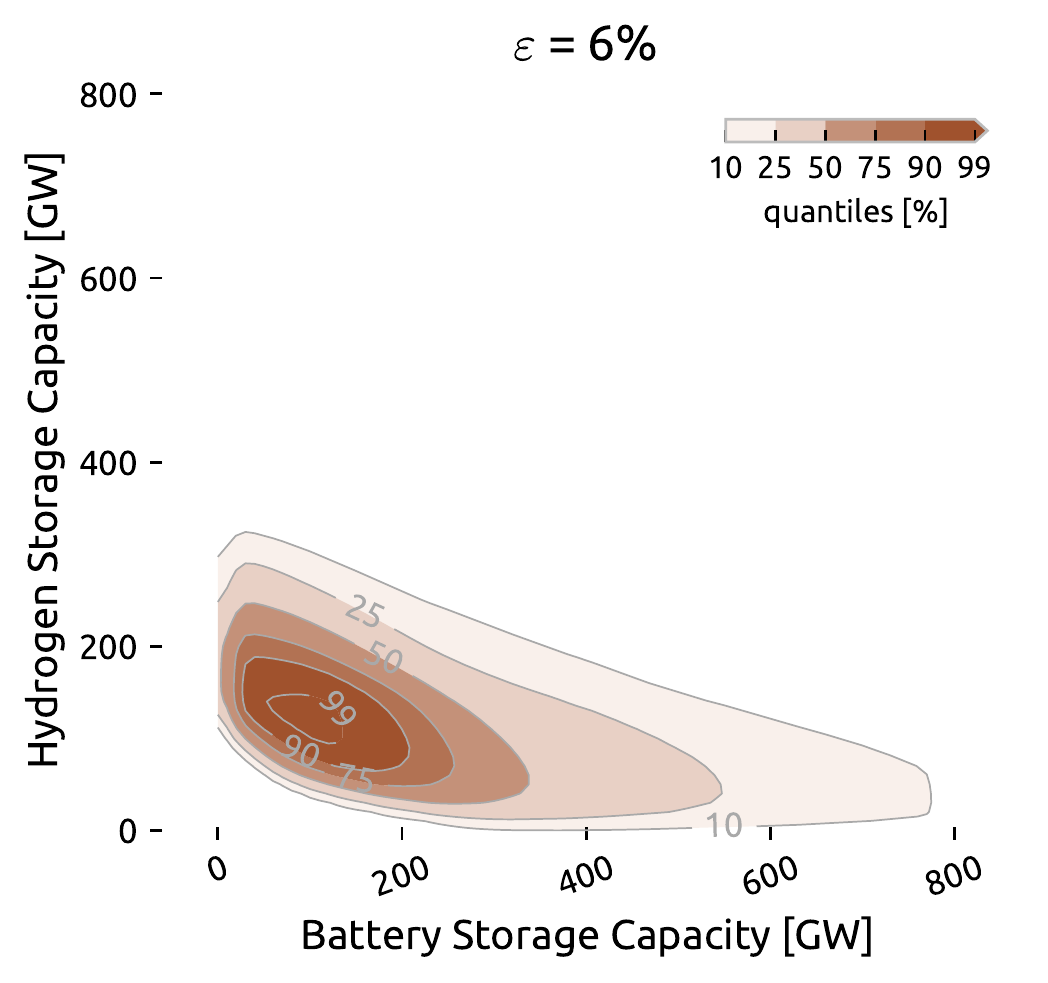}
    \end{subfigure}
    }
    \caption{
      Space of near-optimal solutions by selected pairs of technologies under cost uncertainty.
    Just like in \cref{fig:fuzzycone}, the contour lines depict the overlap of the space of near-optimal alternatives across the parameter space.
    It can be thought of as the cross-section of the probabilistic near-optimal feasible space for a given $\epsilon$
    in two technology dimensions and highlights that the extremes of two technologies from \cref{fig:fuzzycone} cannot be achieved simultaneously.
    }
    \label{fig:dependencies}
\end{figure}

The fuzzy cones from \cref{fig:fuzzycone} look at trade-offs between system cost
and single techologies, assuming that the other technologies can be heavily
optimised. But as there are dependencies between the technologies, in
\cref{fig:dependencies} we furthermore evaluate trade-offs between technologies
for three selected pairs at fixed system cost increase of $\epsilon=6\%$,
addressing which \textit{combinations} of wind and solar
capacity, offshore and onshore turbines, and hydrogen and battery storage are
likely to be cost-efficient.

First, \cref{fig:dependencies:ws} addresses constraints between wind and solar.
The upper right boundary exists because building much of both wind and solar
would be too expensive. The absence of solutions in the bottom left corner
means that building too little of any wind or solar does not suffice to generate
enough electricity. From the shape and contours, we see a high chance that
building 1000 GW of wind \textit{and} 400 GW of solar is within 6\% of the
cost-optimum. On the other hand, building less than 200 GW of solar and 600 GW
of wind is unlikely to yield a low-cost solution. In general, minimising the
capacity of both primal energy sources will shift capacity installations to
high-yield locations even if additional network expansion is necessary and boost
the preference for highly efficient storage technologies. Overall, we can take
away from this that, even considering combinations of wind and solar, a wide
space of low-cost options exists with moderate to high likelihood, although the
range of alternatives is shown to be more constrained.

The trade-off between onshore wind and offshore wind is illustrated in
\cref{fig:dependencies:oo}. Here, the most certain area is characterised by
building more than 600 GW onshore wind, and less than 250 GW offshore wind
capacity. However, there are some solutions with high substitutability between
onshore and offshore wind, shown in the upper left bulge of the contour plot.
Compared to wind and solar, the range of near-optimal solutions is even more
constrained. The key role of energy storage in a fully renewable system is
underlined in \cref{fig:dependencies:hb}. Around 50 GW of each is at least
needed in any case, while highest likelihoods are attained when building 150 GW
of each.

\begin{figure}
    \noindent\makebox[\textwidth]{
    \begin{subfigure}[t]{0.65\textwidth}
     \centering
     \caption{minimal onshore wind with 8\% system cost slack}
     \label{fig:nearviolin:onwind}
     \includegraphics[width=\textwidth, trim=0cm .3cm 3.3cm .63cm, clip]{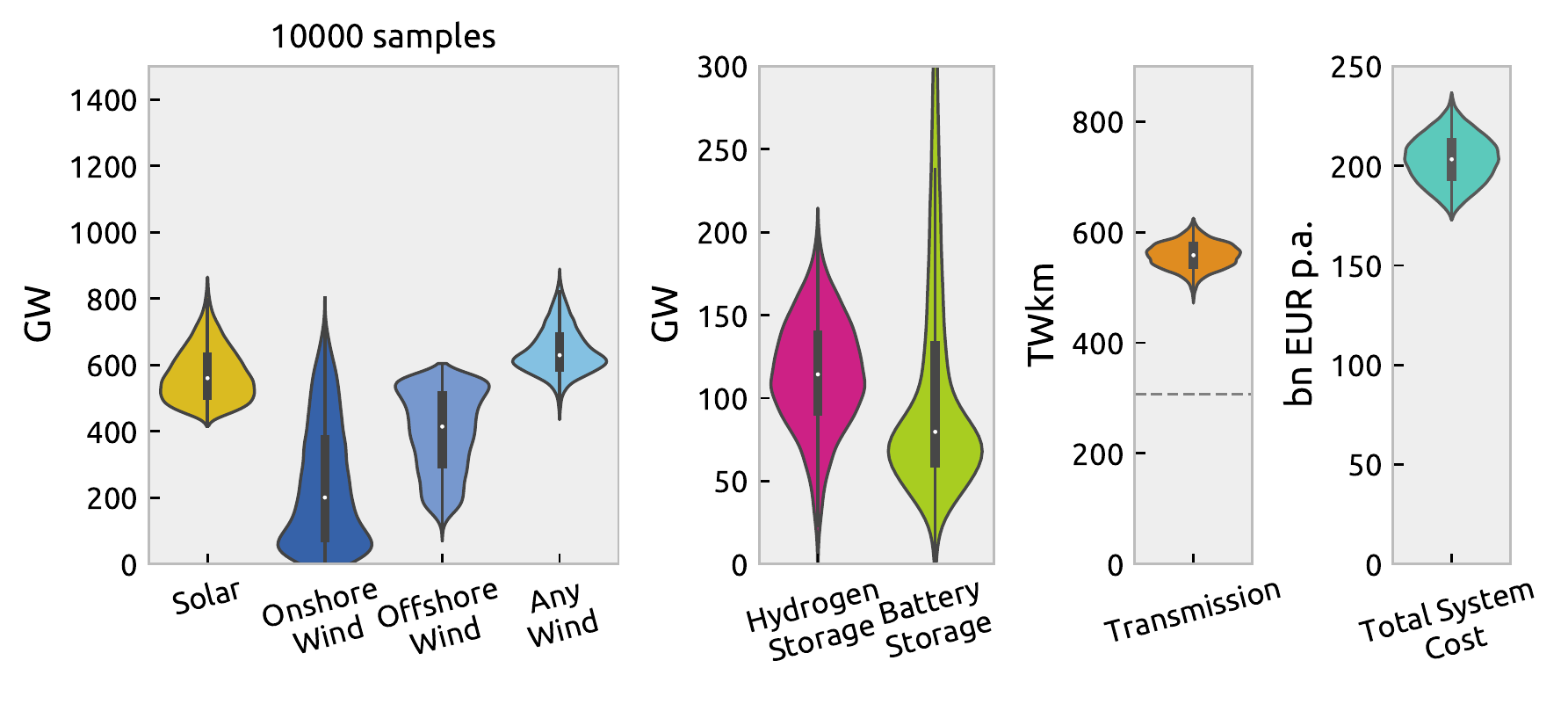}
    \end{subfigure}
    \begin{subfigure}[t]{0.65\textwidth}
     \centering
     \caption{minimal transmission expansion with 8\% system cost slack}
     \label{fig:nearviolin:transmission}
     \includegraphics[width=\textwidth, trim=0cm .3cm 3.3cm  .63cm, clip]{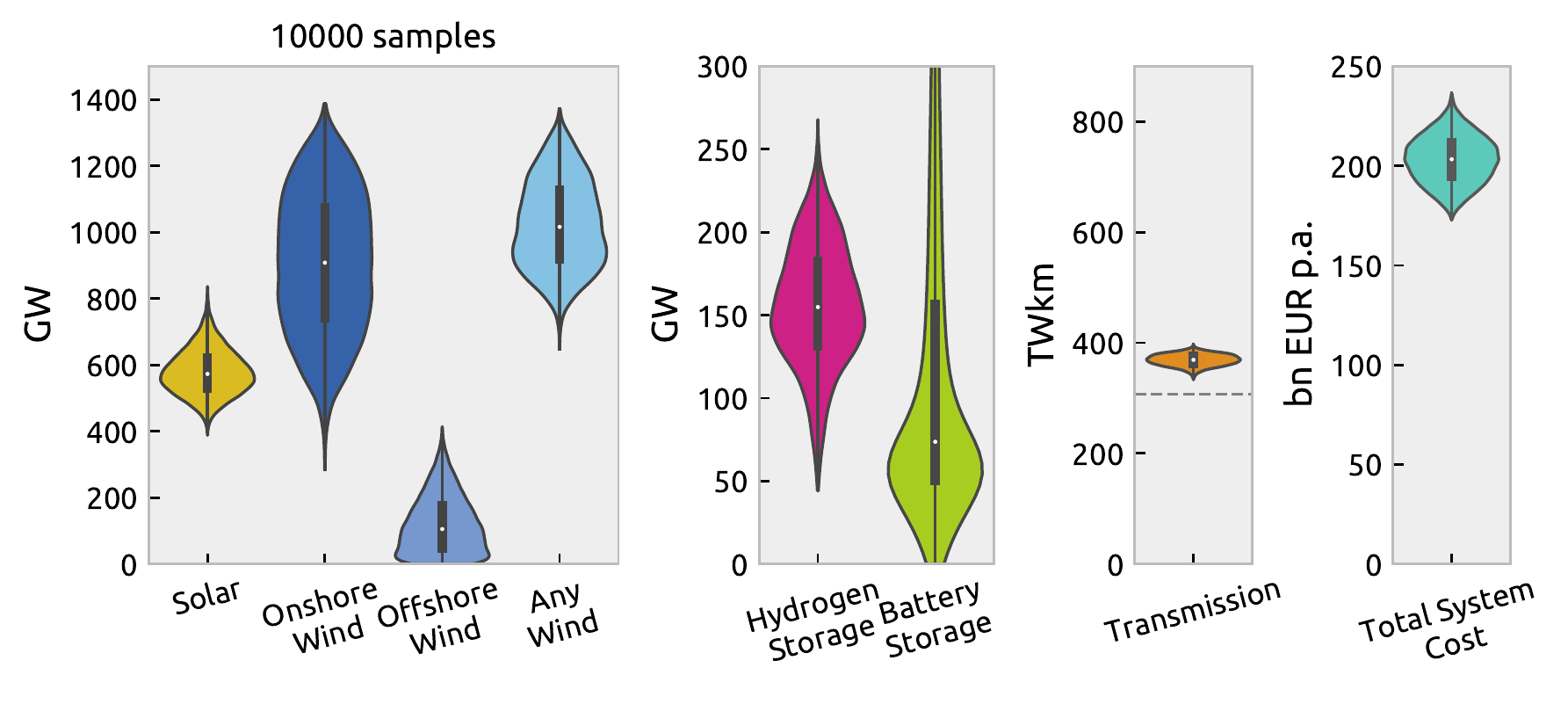}
    \end{subfigure}
    }
    \caption{
      Distribution of total system cost, generation, storage, and transmission capacities
      for two near-optimal search directions with $\epsilon=8\%$ system cost slack.
    }
    \label{fig:nearviolin}
\end{figure}

\subsection{Capacity Distributions at Minimal Onshore Wind and Transmission Grid}

The aforementioned contour plots \crefrange{fig:fuzzycone}{fig:dependencies} outline
what is likely possible within specified cost ranges and subject to technology cost uncertainty,
but do not expose
the changes the overall system layout experiences when reaching for the extremes in one technology.
Therefore, we show in \cref{fig:nearviolin} how the system-wide capacity distributions vary
compared to the least-cost solutions (\cref{fig:violin}) for two exemplary alternative objectives.
For that, we chose minimising onshore wind capacity and transmission expansion
because they are often linked to social acceptance issues.

\cref{fig:nearviolin:onwind} illustrates that reducing onshore wind capacity is
predominantly compensated by increased offshore wind generation but also added solar capacities.
The increased focus on offshore wind also leads to a tendency towards more hydrogen storage,
while transmission expansion levels are similarly distributed as for the least-cost solutions.
From \cref{fig:nearviolin:transmission} we can further extract that avoiding transmission expansion entails
more hydrogen storage that compensates balancing in space with balancing in time,
and more generation capacity overall, where resources are distributed to locations with
high demand but weaker capacity factors and more heavily curtailed.

\subsection{Critical Appraisal}

The need to solve models for many cost projections and near-optimal search
directions in reasonable time means that compromises had to be made in other
modelling dimensions. For instance, the analysis would profit from a richer set
of technologies and further uncertain input parameters, including efficiencies
of fuel cells and electrolysis or the consideration of concentrating solar
power, geothermal energy, biomass, and nuclear to name just a few. But as the
number of considered technologies and parameters rises, so does the
computational burden. Given the already considerable computational efforts
involved in procuring our results, considering the full breadth of technologies
and uncertainties would not have been feasible with the computational resources
available. Moreover, limitations apply to the scope of the analysis which is
limited to the electricity sector does not consider coupling to other energy
sectors. However, accounting for interactions across sectors at high resolution
in similarly set future studies is desirable and in development. Additionally,
we assess no path dependencies via multi-period investments and endogenous
learning, but optimise for an emission reduction in a particular target year
based on annualised costs. We further disregard interannual variations of
weather data by basing the analysis just on a single weather year for computational
reasons. Lastly, aspects such as reserves, system adequacy and inertia have not
been considered.


\section{Conclusion}
\label{sec:conclusion}

In this work, we systematically explore a space of alternatives beyond
least-cost solutions for society and politics to work with. We show how narrowly
following cost-optimal results underplays an immense degree of freedom in
designing future renewable power systems. To back our finding that there is no
unique path to cost-efficiency, we account for the inherent uncertainties
regarding technology cost projections, and draw robust conclusions about the
range of options, boundary conditions and cost sensitivities:

\paragraph{Wide Range of Trade-Offs}
We find that there is a substantial range of options
within 8\% of the least-cost solution
regardless of how cost developments will unfold.
This holds across all technologies individually
and even when considering dependencies between
wind and solar, offshore and onshore wind, as well as hydrogen and battery storage.

\paragraph{Must-Avoid Boundary Conditions}
We also carve out a few boundary conditions which
must be met to keep costs low and are not affected
by the prevailing cost uncertainty.
For a fully renewable power system,
either offshore or onshore wind capacities
in the order of 600 GW
along with some long-term storage technology and
transmission network reinforcement by more than 30\% appears essential.


\paragraph{Technology Cost Sensitivities}
We identify onshore wind cost as the apparent main determinant of system cost,
though it can often be substituted with offshore wind for a small additional cost.
Moreover, the deployment of batteries is the most sensitive to its cost,
whereas required levels of transmission expansion are least affected by cost uncertainty. \\

The robust investment flexibility in shaping a fully renewable power system we
reveal opens the floor to discussions about social trade-offs and navigating
around issues, such as public opposition towards wind turbines or transmission
lines. Rather than modellers making normative choices about how the energy
system should be optimised, we offer methods that present a wide spectrum of
options and trade-offs that are feasible and within a reasonable cost range, to
help society decide how to shape the future of the energy system.



\section*{Acknowledgement}

F.N. and T.B. gratefully acknowledge funding from the Helmholtz
Association under grant no. VH-NG-1352.
The responsibility for the contents lies with the authors.

\section*{CRediT Author Statement}

\textbf{Fabian Neumann:} Conceptualization, Methodology, Investigation, Software, Validation, Formal analysis, Visualization, Writing -- Original Draft, Writing -- Review \& Editing
\textbf{Tom Brown:} Conceptualization, Writing -- Review \& Editing, Supervision, Project administration, Funding acquisition

\section*{Data Availability}

The code to reproduce the experiments is available at \href{https://github.com/fneum/broad-ranges}{github.com/fneum/broad-ranges}.
We also refer to the documentation of PyPSA (\href{https://pypsa.readthedocs.io}{pypsa.readthedocs.io}) and
PyPSA-Eur (\href{https://pypsa-eur.readthedocs.io}{pypsa-eur.readthedocs.io}).

\addcontentsline{toc}{section}{References}
\renewcommand{\ttdefault}{\sfdefault}
\bibliography{library}


\end{document}